\documentclass[aps,pre,twocolumn,groupedaddress,amsmath,showpacs,floatfix]{revtex4}

\usepackage{graphicx,subfigure,feynmf,bm}

\begin{document}


\title{Random Isotropic Structures and Possible Glass Transitions \\ in Diblock Copolymer Melts}


\author{Cheng-Zhong Zhang}
\affiliation{Division of Chemistry and Chemical Engineering, California Institute of Technology, Pasadena, California 91125, USA}

\author{Zhen-Gang Wang}
\affiliation{Division of Chemistry and Chemical Engineering, California Institute of Technology, Pasadena, California 91125, USA}

\date{\today}

\begin{abstract}
We study the microstructural glass transition in diblock-copolymer melts using a thermodynamic replica approach. 
Our approach performs an expansion in terms of the natural smallness parameter--the inverse of the scaled degree of polymerization $\bar{N}$--which allows us to systematically study the approach to mean-field behavior as the degree of polymerization increases. 
We find that in the limit of infinite chain length, both the onset of glassiness and the vitrification transition (Kauzmann temperature) collapse to the mean-field spinodal, suggesting that the spinodal can be regarded as the mean-field signature for glass transitions in this class of  microphase-separating system. 
We also study the order-disorder transition (ODT) within the same theoretical framework; in particular, we include the leading-order fluctuation corrections due to the cubic interaction in the coarse-grained Hamiltonian, which has been ignored in previous studies of the ODT in block copolymers.  
We find that the cubic term stabilizes both the ordered (body-centered-cubic) phase and the glassy state relative to the disordered phase. In melts of symmetric copolymers the glass transition always occurs after the order-disorder transition (below the ODT temperature), but for asymmetric copolymers, it is possible for the glass transition to precede the ordering transition.
\end{abstract}


\maketitle

\section{\label{sec:1}Introduction}

Block copolymers are macromolecules built with blocks of chemically distinct monomers. 
Melts of block copolymers are attractive from both theoretical and experimental standpoints, as they undergo microphase transitions and produce diverse ordered microstructures \cite{Hamley98,Hamley04,BatesFredricksonRev91,BatesFredricksonPhysToday}.

The simplest block copolymer is the $AB$ diblock copolymer made of two types of monomers $A$ and $B$.  
Below the order-disorder transition (ODT) temperature, a diblock-copolymer melt can exhibit rich mesophases \cite{Bates94a}, including body-centered-cubic (bcc), hexagonally ordered cylinder (hex), lamellar (lam), and several bicontinuous (e.g., gyroid) structures. 
Experimentally these structures have been identified using transmission electron microscopy (TEM) \cite{Bates82, ELThomas94Mac}, small-angle neutron scattering (SANS) \cite{Bates82,Bates90a,Bates92a}, and dynamic mechanical measurements \cite{Bates90a,Bates90c,Bates92a}. 
Theoretically these structures are well described by the self-consistent--mean-field theory \cite{MatsenSchick,MatsenBatesRev96}.

Generally these periodically ordered structures are expected to be the thermodynamically equilibrium states \cite{Bates90a}. 
However, they are difficult to attain either in experiments \cite{Bates90a} or in computer simulations \cite{Fraaije96a}. 
Bates and co-workers \cite{BatesFredricksonRev91,Bates90a} found that quenching a nearly symmetric diblock-copolymer melt without symmetry-breaking external field, such as reciprocal shearing, generally results in isotropic, locally microphase-separated structures with a characteristic length scale of the radius of gyration of the polymer. In addition, such structures were also obtained (as a rule) in dynamic-density-functional calculations \cite{Fraaije96a, Fraaije97b,Hamley98}. 
The ordering kinetics in these random structures are very slow, suggesting that they are metastable states corresponding to free-energy minima.
It is therefore quite possible that the ordered phases, though energetically favored, are not easily reached due to the kinetic trapping caused by the presence of a large number of metastable free-energy minima. These metastable states correspond to the locally microphase-separated states without long-range order.

Dynamic mechanical measurements by Bates and co-workers on melts of both symmetric, lamellae-forming, and asymmetric, hex-forming PE--PEP [partially deuterated poly(ethylene-propylene)--poly(ethylethylene)] copolymers revealed \cite{Bates90a,Bates92a} that the system may be frozen in random structures upon a deep quench. 
Comparing the quenched sample with the slowly-supercooled sample and the shear-ordered sample, they found that the quenched sample exhibits very slow relaxations and extraordinarily large elastic moduli at low frequencies; but the supercooled sample behaves more like the disordered melt continuously extended to below the ODT temperature. 
Balsara and co-workers \cite{Balsara98a,Balsara01a} studied the grain structure of asymmetric, hex-forming PI-PS (polyisoprene--polystyrene) melt by light scattering, SANS, and rheological measurements. Similar to the findings of Bates \textit{et~al.}, they found that upon a deep quench, randomly microphase-separated structures are obtained, which do not appear to evolve towards the equilibrium structure with long range order within the time scales of the experiments. 
Besides these, Pochan \textit{et al.} \cite{Gido96Mac} found randomly oriented wormlike cylinder structures in an I$_2$S [polyisoprene(I)-polystyrene(S)] star copolymer system,. 

The above results suggest that the ordering process in block-copolymer melts follow a two-step mechanism: 
a fast step in which unlike monomers locally phase separate into random, macroscopically isotropic structures with domains of the size of a single polymer, followed by a domain coarsening (or growth) step in which local defects in the random microstructures anhilate and long range order is developed. 
The second step is generally much slower than the first and most likely involves activated processes. 
Therefore a rapid deep quench can result in randomly microphase-separated structures that are kinetically trapped and unable to develop long range order within normal laboratory time scales.

This two-step mechanism is in fact consistent with the thermodynamic two-step scenario implicit in the Fredrickson-Helfand (FH) fluctuation theory for diblock-copolymer melts \cite{FredricksonHelfand} (which only applies to symmetric or nearly symmetric copolymers). 
Instead of the featureless background as assumed in the random-field-approximated structure factor of Leibler \cite{Leibler}, the FH theory suggests that when the temperature approaches the ODT, the disordered state is a fluctuating, heterogeneous structure consisting of locally $A$- and $B$-rich domains, which then orders into periodic mesophases upon further cooling. 

The most dramatic manifestation of the first step is the existence of disordered-spherical-micelle state in highly asymmetric copolymer melts, which almost has the appearance of a distinct phase between the featureless disordered phase and the bcc-ordered phase \cite{Register94,Register96,Schwab96,Kim99Mac,Han00a,Hashimoto03a,Han03a,Lodge02}. 
The micelle state was first predicted by Semenov \cite{Semenov89Mac}. Recently Dormidontova and Lodge \cite{Lodge01} extended the Semenov theory by including the translational entropy of the disordered spherical micelles and predicted a phase diagram that is in qualitative agreement with experiments. 
More recently, Wang and co-workers \cite{Wang05Mac} examined the nature of the disordered spherical micelles and their connection to concentration fluctuations using the self-consistent-field theory. 
Taking a nucleation perspective, these authors showed that the disordered micelles are large, localized concentration fluctuations through a thermally activated process.

In this work, we study the metastable states consisting of random structures in block-copolymer melts and address the possibility of glass transition using a thermodynamic replica approach. 
This approach was first proposed by Monasson \cite{Monasson95} and subsequently employed by a number of authors in studying structural glass transitions \cite{Mezard2000a,Parisi2000a,Schmalian2000a,Westfahl2001,Wolynes2004}. 
In this framework, the onset of glassiness is identified with broken ergodicity \cite{Palmer}, which occurs as a result of the appearance of an exponentially large number of metastable free-energy minima \cite{Nussinov1999}. The broken ergodicity is manifested through a nonvanishing long-time correlation (here manifested as the cross replica correlation function), whose first appearance defines the onset temperature of glassiness $T_A$ (also called the dynamic glass transition temperature \cite{Westfahl2001, Monasson95}). 
An equivalent Kauzmann temperature $T_K$ as in molecular liquids \cite{Debenedetti00} can also be defined as signaling the complete vitrification of the random structures.

The possibility of glass transitions in bicontinuous microemulsions--a system closely related to diblock copolymers--was recently examined by Wu, Westfahl, Schmalian and Wolynes \cite{Wolynes2002b}, using both a dynamic mode-coupling theory and the thermodynamic replica approach. 
There authors have also studied glass transitions in the Coulomb-frustrated-magnet model using the replica method with a self-consistent-screening approximation \cite{Schmalian2000a, Westfahl2001} and, more recently, a local-field calculation \cite{Wolynes2004}. 
Both the microemulsion and the Coulomb-frustrated-magnet systems belong to the general class of models first proposed by Brazovskii \cite{Brazovskii75}, featuring the existence of low-energy excitations around some finite wave number $q_m$ and the formation of microphase-separated structures with length scales $\sim 1/q_m$ at low temperatures. 
These studies showed that as a result of the large degeneracy in ground states \cite{Nussinov1999}, a glass transition can occur when the ratio of the correlation length of the system to the modulation length $2 \pi /q_m$ exceeds some critical value. 
Similar conclusions were also obtained by Grousson \textit{et~al}. \cite{Grousson2002a} using the mode-coupling theory.

Our work follows a similar approach to that employed by Schmalian and co-workers \cite{Schmalian2000a, Westfahl2001}.  
However, we perform calculations specifically for the block-copolymer system by taking advantage of the natural smallness parameter (the inverse of the scaled degree of polymerization, $\bar{N}$); this allows us to study how the glass transitions are affected by increasing the chain length of the polymer when the system gradually approaches the mean-field limit.  
An important conclusion of our work is that in the limit of infinitely long chains, both the onset of glassiness and the Kauzmann temperature coincide with the mean-field spinodal of the disordered phase. Therefore the spinodal is the mean-field signature for the glass transition in the block copolymer system; the same conclusion is likely to hold in general for microphase-separating systems. 
Another feature of our work is the inclusion of the order-disorder transition in the phase diagram. This is important because it places the glass transition in proper relationship to the ordering transition.
We find that, for symmetric, lam-forming copolymers, the glass-transition temperatures are below the ODT temperature, while for asymmetric, sphere-forming copolymers, the onset of glassiness can precede the ODT into the bcc phase. 
On a technical point, we propose a method for incorporating fluctuations due to the cubic interaction in the Brazovskii model, using a renormalization scheme motivated by the $1/n$ expansion of the $n$-vector model in critical phenomena.  The effects of these  fluctuations have not been addressed in any of the previous studies \cite{FredricksonHelfand,DelaCruz91,DobryninErukhimovich,BarratFredrickson, FredricksonBinder} on block-copolymer systems. 
We find that in the leading-order approximation these fluctuations stabilize both the bcc phase and the glassy state.

\section{\label{sec:2}Model and Solution}

\subsection{\label{sec:2.A}Model description}

We consider the melt of $AB$ diblock copolymers of degree of polymerization $N=N_A+N_B$ and block composition $f=N_A/N$. 
The monomer volume $v$ and Kuhn length $b$ are taken to be equal for both monomers. 
We describe the thermodynamics of the system using the random-field-approximated (RPA) free-energy functional with local approximations for the cubic and quartic interactions \cite{Leibler,OhtaKawasaki,FredricksonHelfand} as the Hamiltonian
\begin{eqnarray}
\mathcal{H}[\phi] = & &\frac{1}{Nv} \left[\frac{1}{2} \int\frac{d^{3}q}{(2\pi)^{3}}\phi(-q)\gamma_{2}(q,-q)\phi(q)\right.\nonumber\\
                    & &\left.+\frac{\gamma_{3}}{3!}\int d^{3}x\phi(x)^{3}+\frac{\gamma_{4}}{4!}\int d^{3}x\phi(x)^{4}\right],\label{eq:2.2}
\end{eqnarray} 
where the order parameter $\phi \equiv \rho_A(x)v - f$ is the density deviation from the mean value. Throughout the paper we take $k_BT=1$ except for our discussion of the thermodynamic approach to the glass transition in Sec.~\ref{sec:2.C}. 
To simplify the notation, we use plain letters ($x$, $q$, etc.) to denote position and wave vectors; when the plain letter is used to denote the magnitude of the wave vector (wave number), the context should make it clear.

Near the mean-field spinodal $\gamma_{2}(q,-q)$ can be approximated as
\[\gamma_2(q,-q)=\frac{c^2}{4}\left(q^{2}Nb^2-q_m^2Nb^2\right)^{2}+2\left(\chi N\right)_S-2\chi N,\]
where $\chi N$ is the Flory-Huggins interaction parameter between $A$ and $B$ blocks, $(\chi N)_{S}$ is its value at the spinodal, and $c$ is a parameter independent of $N$. 
$(\chi N)_S$, $c$, and $q_m$ are functions of $f$ and $N$, which can be calculated using the RPA theory of Leibler \cite{Leibler}. 
Note that Eq.~(\ref{eq:2.2}) as a Hamiltonian is applicable to a broad class of copolymer systems, including multiblock copolymers \cite{DelaCruz91} and copolymer/homopolymer blends \cite{Muthukumar97}, where the dependence on chain architectures, block compositions, and volume fractions of copolymers can be incorporated into the parameters $(\chi N)_S$, $q_m$, etc. Therefore our results on diblock-copolymer systems should be qualitatively applicable to these systems as well.

The degree of polymerization, $N$, plays the role of Ginzburg parameter, which controls the magnitude of fluctuations \cite{FredricksonHelfand}. 
To highlight this feature, we nondimensionalize the lengths and wave numbers by the ideal end-to-end distance of the polymer: $\bar{x} \equiv x/(\sqrt{N}b)$, $\bar{q}\equiv q \sqrt{N}b$, $\bar{q}_m \equiv q_m \sqrt{N}b$, and concurrently rescale the order parameter as $\bar{\phi}(\bar{x}) \equiv \phi(x) c\bar{q}_m$, $\bar{\phi}(\bar{q}) \equiv \phi(q) c\bar{q}_m /(\sqrt{N}b)^3$.  
Now the Hamiltonian [Eq.~(\ref{eq:2.2})] becomes
\begin{eqnarray}
\mathcal{H}[\phi] &  = & \frac{\sqrt{N}b^{3}}{v}\left[\frac{1}{2}\int\frac{d^3{\bar{q}}}{(2\pi)^{3}}g(\bar{q})^{-1}\bar{\phi}(\bar{q})\bar{\phi}(-\bar{q})\right.\nonumber\\                             && +\left.\frac{\eta}{3!}\int d^{3}\bar{x}\bar{\phi}(\bar{x})^{3}+\frac{\lambda}{4!}\int d^{3}\bar{x}\bar{\phi}(\bar{x})^{4}\right]\nonumber\\                                    & = &\bar{N}^{1/2}H[\bar{\phi}]\label{eq:2.11},
\end{eqnarray}
where
\begin{eqnarray}
g(\bar{q})^{-1} & = & \frac{1}{4\bar{q}_m^2}\left(\bar{q}^{2}-\bar{q}_{m}^{2}\right)^{2}+\tau_{0}\bar{q}_m^2,\label{eq:2.12}\\
\tau_{0} & = & \frac{2\left(\chi N\right)_S-2\chi N}{c^{2}\bar{q}_m^{4}},\label{eq:2.13}\\
\eta & = & \frac{\gamma_3}{c^3\bar{q}_m^3},\\
\lambda & = & \frac{\gamma_4}{c^4\bar{q}_m^4}.
\end{eqnarray}
The scaled couplings $\eta$ and $\lambda$ are, respectively, the same as $N\Gamma_3$ and $N\Gamma_4$ defined in Ref.~\cite{FredricksonHelfand}.  
For notational simplicity, we drop the overbars on the variables and the order parameter henceforth.

We point out that, although the parameters in Eq.~(\ref{eq:2.11}) are written in molecular terms, this model is best interpreted as phenomenological.
The random-phase approximation used in deriving the Hamiltonian, the approximation of higher-order interactions as spatially local, and the truncation at quartic order in the order-parameter expansion--all introduce inaccuracies whose effects are difficult to evaluate \cite{Stepanow03}.  
In particular, the order-parameter expansion to quartic order is not justified for strongly asymmetric block compositions as chain stretching effects become important and the weak-segregation assumption no longer holds \cite{Bates90b}.  However, we note that taking Eq.~(\ref{eq:2.11}) as the Hamiltonian, one can reproduce the experimental phase diagram of microphase transitions qualitatively at all compositions, including the disordered spherical-micelle states at very asymmetric compositions, as the state-of-the-art self-consistent-field theory.   
Therefore while the quantitative accuracy of our theory may not be reliable, we expect that most of our predictions should be qualitatively correct.  Such an expectation is further boosted by the general success of the Fredrickson-Helfand theory [also using Eq.~(\ref{eq:2.11}) as the Hamiltonian] in capturing many key features of the physics of diblock-copolymer melts at length scales comparable to or larger than the size of the polymer chain. 

In addition, studying glass transitions in the system described by Eq.~(\ref{eq:2.11}) is of intrinsic theoretical value, as Eq.~(\ref{eq:2.11}) corresponds to the weak-coupling limit of the Brazovskii model. Therefore our results elucidate the physics of systems in the Brazovskii class in this limit.

Finally we notice that the parameter $\bar{N}^{1/2} \equiv N^{1/2} b^3/v$ (henceforth referred to as the ``chain length'') is a natural combination emerging in any study of the fluctuation effects in polymer melts, which gives the number of other chains within the spatial extension of a single polymer chain \cite{FredricksonHelfand, Wang02a}. 
$\bar{N}$ plays a role similar to $1/\hbar$ in quantum field theory \cite{Jackiw}--controlling the magnitude of fluctuations. In the limit of $\bar{N}\rightarrow\infty$, mean-field behavior is recovered. For systems with large but finite $\bar{N}$ we can apply a systematic loop expansion using $1/\bar{N}^{1/2}$ as the smallness parameter. 

The presence of the $\bar{N}^{1/2}$ factor in front of the Hamiltonian also has important consequences on the free-energy barriers separating the multiplicity of free-energy minima.  In the mean-field approximation, we expect that the free-energy barriers should be proportional to this factor.  For long polymer chains, the barriers can be much larger than the thermal energy, resulting in slow relaxations from these free energy minima to the lower-free energy ordered phases and between the metastable minima themselves. This justifies the application of the energy-landscape theory of glass transitions in polymer systems.

\subsection{\label{sec:2.B}Ordered states and order-disorder transition}

Our current understanding of the effects of fluctuations on the ODT in block-copolymer melts is largely based on the Brazovskii-Leibler-Fredrickson-Helfand (BLFH) theory \cite{Brazovskii75, Leibler, FredricksonHelfand}. 
This theory uses the self-consistent Brazovskii approximation (a Hartree-type approximation) for the quartic interaction and ignores fluctuations due to the cubic interaction. Therefore, strictly speaking, it is only valid for symmetric or nearly symmetric block copolymers where cubic interaction is small (see our discussions at the end of this subsection).
Here we extend this theory to include the leading-order one-loop correction from the cubic interaction, which accounts for the fluctuation effects due to asymmetry in the copolymer composition. 
This improved theory should give more accurate predictions on the ODT in asymmetric copolymer melts (and other asymmetric systems) and, more important, enables a consistent comparison with the glass transition in the same system, where the cubic term is shown to play a dominant role.

As in previous weak-segregation theories \cite{Leibler,FredricksonHelfand}, we adopt the single-mode approximation for the periodic microphases, representing the density wave by
\begin{equation}
\varphi(x) = a\sum_{j}\left[\exp(iQ_{j}\cdot x)+\exp(-iQ_{j}\cdot x)\right],\label{eq:3.1}
\end{equation}
where $a$ is the magnitude of the density wave and $Q_{j}(1\leq j\leq n)$ are the first set of vectors on the reciprocal lattice of the periodic structure of the ordered microphases \cite{Leibler,FredricksonHelfand}. Now we introduce the fluctuation field around the minimum, $\psi(x)=\phi(x)-\varphi(x)$, and perform an expansion of the Hamiltonian $H[\phi]$ in Eq.~(\ref{eq:2.11}) around $\varphi$. The fluctuation part of $H[\phi]$ is
\begin{widetext}
\begin{eqnarray}
\Delta H[\psi; \varphi] && = H[\psi+\varphi]- H[\varphi]\nonumber\\
                           & &= \frac{1}{2}\int\frac{d^3q}{(2\pi)^3}\psi(-q)g(q)^{-1}\psi(q)+\frac{\eta}{3!}\int d^3x\psi(x)^3+\frac{\lambda}{4!}\int d^3x\psi(x)^4\nonumber\\                       &  &+\frac{\eta}{2}\int\frac{d^{3}q_{1}d^3q_{2}d^3q_{3}}{(2\pi)^9}\psi(q_1)\psi(q_2)\varphi(q_3)\delta^{3}(q_1+q_2+q_3)\nonumber\\ 
                          &&+\frac{\lambda}{4!}\int\frac{d^{3}p_{1}d^3p_{2}d^{3}p_{3}d^3p_{4}}{(2\pi)^{12}}\left[4\psi(p_1)\psi(p_2)\psi(p_3)\varphi(p_4)+6\varphi(p_1)                                    \varphi(p_2)\psi(p_3)\psi(p_4)\right]\delta^{3}(p_1+p_2+p_3+p_4).\label{eq:3.6}
\end{eqnarray}
The linear term of $\psi$ vanishes because $\varphi$ is at the minimum of the Hamiltonian.  
For the quadratic term we only keep the dominant isotropic part 
\begin{equation}
D(q)^{-1}=g(q)^{-1}+n\lambda a^2,\label{eq:3.2}
\end{equation}
which is defined as the shifted bare propagator.

The free energy (effective potential) of the microphase-separated system is given by
\begin{eqnarray}
F[\varphi] & = &-\bar{N}^{-1/2}\ln\left\langle\exp\left\{-\bar{N}^{1/2}{ H[\phi]}\right\}\right\rangle\nonumber\\
                & = & H[\varphi]-\bar{N}^{-1/2}\ln\left\langle\exp\left\{-\bar{N}^{1/2} \Delta H[\psi;\varphi]\right\}\right\rangle.\label{eq:3.7}
\end{eqnarray}
Here the free energy is scaled by $\bar{N}^{-1/2}$ such that the mean-field part $H[\varphi]$ is independent of $\bar{N}$ and reduces to the Leibler free energy \cite{Leibler}.
The second term in Eq.~(\ref{eq:3.7}) contains corrections due to the fluctuation part of the Hamiltonian [Eq.~(\ref{eq:3.6})]. In the one-loop approximation we have
\begin{equation}
F[\varphi]=H[\varphi]+\frac{1}{2\bar{N}^{1/2}}\mathtt{Tr}\ln\mathcal{G}_H^{-1}-\frac{\lambda}{8\bar{N}}\left[\int\frac{d^3{q}}{(2\pi)^{3}}\mathcal{G}_H(q)\right]^{2}-\frac{\eta^{2}}{12\bar{N}}\int\frac{d^{3}pd^{3}q}{(2\pi)^{6}}\mathcal{G}_H(p)\mathcal{G}_H(q)\mathcal{G}_H(-q-p),\label{eq:3.4}
\end{equation}
where $\mathcal{G}_H(q)$ is the Hartree-renormalized propagator determined from
\begin{equation}
\mathcal{G}_H(q)^{-1}=D(q)^{-1}+\frac{\lambda}{2\bar{N}^{1/2}}\int\frac{d^3{k}}{(2\pi)^{3}}\mathcal{G}_H(k).\label{eq:3.3}
\end{equation}
Our one-loop approximation is slightly different from the conventional diagrammatic expansion; details are discussed in Appendix \ref{Appendix:C}. 

Under this approximation the renormalized correlation function is given by
\begin{eqnarray}
\mathcal{G}(q)^{-1}&=&\left\{\frac{\delta^2F[\varphi]}{\delta\varphi(q)\delta\varphi(-q)}\right\}^{-1}\nonumber\\
              &=&\mathcal{G}_H(q)^{-1}-\frac{\eta^{2}}{2\bar{N}^{1/2}}\int\frac{d^{3}k}{(2\pi)^{3}}\mathcal{G}_H(k)\mathcal{G}_H(q-k).\label{eq:3.5}
\end{eqnarray}
\end{widetext}
In the replica calculation $\mathcal{G}(q)$ gives the renormalized diagonal correlation function in the replica space.

The second term in Eq.~(\ref{eq:3.5}), corresponding to the one-loop cubic diagram, was absent in previous theories on the ODT, as it was shown to be subdominant to the Hartree term [the first term in Eq.~(\ref{eq:3.5})] near the mean-field ODT for asymmetric copolymers \cite{Brazovskii75,SwiftHohenberg77}.  
The arguments for ignoring this term no longer hold for the supercooled disordered phase (below the ODT temperature), and here we need a free-energy function that remains valid even below the mean-field spinodal temperature. Therefore the one-loop cubic term cannot be dropped as in Refs.~\cite{Brazovskii75} and \cite{FredricksonHelfand}.
Also in the $1/\bar{N}$ expansion employed here (equivalent to a loop expansion), the one-loop cubic term is of the same order as the Hartree term and their numerical values are comparable in the part of the phase diagram of interest, except for nearly symmetric compositions when the cubic term is small \footnote{In a complementary perturbative expansion motivated by the $1/n$ expansion, this cubic diagram is subdominant to the Hatree term (see Appendix~\ref{Appendix:C} for a discussion). But their numerical values turn out to be comparable.}. 
Furthermore, as shown in Appendix~\ref{Appendix:A} the corresponding term is the leading term in the self-consistent equation for the cross-replica correlation function.  Earlier work also showed that it is the leading term that generates long-time correlations in the mode-coupling theory for glass transitions \cite{KirkpatrickJPhysA89}.
We therefore include the one-loop cubic term in our treatment of the ODT to have a consistent comparison with glass transition.

\subsection{\label{sec:2.C}Random structures and glass transition}

Traditionally two different approaches have been developed to study frustrated systems with quenched disorder. 
The dynamic approach, most notably the mode-coupling theory \cite{Gotze}, focuses on dynamic correlation functions (e.g., the Edwards-Anderson order parameter defined as the long-time spin-spin correlation function in the Ising-spin-glass model \cite{EdwardsAndersonII}) and characterizes the glassy state with nonvanishing long-time correlations and broken ergodicity. 
On the other hand, the equilibrium thermodynamic approach, including the density functional approach \cite{Wolynes1985} and the replica approach \cite{MezardParisibook}, describes glass transitions in terms of the energy-landscape features of the system \cite{Debenedetti00}.  
The connection between these two approaches was explicitly demonstrated in the mean-field spin-glass models \cite{KirkpatrickPRA87a, MezardParisibook} where it was shown that these two approaches yield consistent predictions.  We now briefly describe the essential concepts in the thermodynamic approach.

The central assumption in the thermodynamic approach is that the dynamic behavior of glass-forming systems reflects the underlying free-energy-landscape features \cite{Debenedetti00, AdamGibbs, KirkpatrickPRA89a, Monasson95, Parisi2000a}.  
At high temperatures, there is only one minimum corresponding to the uniform liquid state.  
As temperature decreases, multiple metastable minima begin to appear that are separated by sizable activation barriers, and below some temperature $T_A$, the number of these minima becomes thermodynamically large, giving a finite contribution to the partition sum of these ``disjoint'' metastable states and generating extensive configurational complexity manifested in a nonvanishing configurational entropy \cite{KirkpatrickPRA89a}. 
This signals the onset of glassiness or broken ergodicity \cite{Palmer} in the sense that within times scales of typical liquid relaxations, the system is trapped in these metastable free-energy minima; transitions between the minima, however, can still occur through activated processes \cite{KirkpatrickPRA87a}. 
Dynamically, one expects a significant slowing down of structural relaxations, often accompanied by the appearance of long plateaus in the time correlation functions \cite{Debenedetti00, KirkpatrickJPhysA89}. 
Complete vitrification occurs at a lower temperature $T_K$, below which the system is dominated by one or less than an exponentially large number of deep free energy minima; thermodynamically, this is signaled by the vanishing of the configurational entropy.  $T_K$ is often termed the ideal glass transition temperature and is conceptually identified with the underlying thermodynamic glass transition at which the viscosity of the supercooled liquid diverges \cite{AdamGibbs, Debenedetti00, Monasson95, Kauzmann}.

Recently Monasson \cite{Monasson95} proposed a replica method which allows explicit implementation of the thermodynamic approach for studying structural glasses resulting from self-generated randomness. Using this method Westfahl \textit{et~al.} \cite{Westfahl2001} successfully predicted the glass transitions in the Coulomb-frustrated-magnet model. 
Here we adopt this approach to study the glass transition in block-copolymer melts. 

Following \cite{Monasson95}, we introduce an external pinning field $\zeta$ and calculate the pinned free energy of the system, $F[\zeta]$
\begin{widetext}
\begin{equation}
F[\zeta]=-k_BT\ln Z[\zeta]=-k_BT\ln\int\mathcal{D}\phi\exp\left(-\frac{1}{k_BT}\left\{\mathcal{H}[\phi]+\frac{\alpha}{2}\int d^{3}x\left[\phi(x)-\zeta(x)\right]^{2}\right\}\right), \label{eq:2.4}
\end{equation}
where $\alpha>0$ is the coupling between the pinning field and the order parameter. [In Eqs.~(\ref{eq:2.4})--(\ref{eq:2.6}), we reintroduce the $k_BT$ factor in order to allow explicit temperature derivatives.]
The effect of $\zeta$ is to locate the basins on the free-energy landscape.  
The coupling constant will be taken to be infinitesimally small at the end and serves as a convenient device for breaking ergodicity--localizing the system into separate basins. 
Its role is similar to that of the infinitesimal field that breaks the up-down symmetry of the Ising model below the critical temperature.  
One can show that the minima of $F[\zeta]$ coincide with those of the effective potential of $\mathcal{H}[\phi]$ as $\alpha\rightarrow 0$; proof is given in Appendix~\ref{Appendix:D}.
Thus $\zeta$ serves as a running index for labeling different basins on the free-energy landscape, and sampling the configuration space of $\zeta$ gives information on the metastable free-energy minima (the energy minima with their location fluctuations) of the system.  Therefore one can use $F(\zeta)$ as an ``effective Hamiltonian" for the metastable free energy minima and compute the ``quenched-average'' free energy 
\begin{equation}
\bar{F}=\frac{\int\mathcal{D}\zeta F[\zeta]\exp\left\{-F[\zeta]/k_BT\right\}}{\int\mathcal{D}\zeta\exp\left\{-F[\zeta]/k_BT\right\}}.
\end{equation} 

If the system is fully ergodic, one can verify that $\bar{F}$ is equal to the equilibrium free energy 
\[F=-k_BT\ln\int\mathcal{D}\phi\exp\{-\mathcal{H}[\phi]/k_BT\}\] 
in the thermodynamic limit as $\alpha\rightarrow 0^+$. However, when ergodicity is broken $\lim_{\alpha\rightarrow0^+}\bar{F}$ can be different from $F$. Their difference 
\begin{equation}
\bar{F}-F=TS_c
\end{equation}
defines the configurational entropy that measures the configurational complexity due to an exponentionally large number of metastable states \cite{Monasson95, Westfahl2001, MezardParisibook, Palmer}. 
In the thermodynamic approach, $S_c$ jumps discontinuously from zero to an extensive finite value at $T_A$, implying broken ergodicity due to disjoint metastable states; $S_c$ decreases upon further cooling and vanishes at $T_K$, when the system becomes completely vitrified.

To calculate $S_c$ it is convenient to introduce the ``replicated'' free energy
\begin{equation}
F_{m}=-\lim_{\alpha\rightarrow0^{+}}\frac{k_BT}{m}\ln\int\mathcal{D}\zeta\exp\left\{-\frac{m}{k_BT}F[\zeta]\right\}
         =-\lim_{\alpha\rightarrow0^{+}}\frac{k_BT}{m}\ln\int\mathcal{D}{\zeta}Z[\zeta]^{m}=-\frac{k_BT}{m}\ln Z_{m}, \label{eq:2.5}
\end{equation}
where $T/m$ is introduced as the effective temperature conjugate to $F[\zeta]$. $\bar{F}$ and $S_c$ are obtained from Eq.~(\ref{eq:2.5}) straighforwardly as
\begin{eqnarray}
\bar{F}&=&\left.\frac{\partial(mF_{m})}{\partial m}\right|_{m=1},\\
S_c &=& -\left.\frac{\partial F_{m}}{\partial (T/m)}\right|_{m=1}=\left.\frac{1}{T}\frac{\partial F_{m}}{\partial m}\right|_{m=1}.\label{eq:2.7}
\end{eqnarray}
When $m$ is an integer, $Z_m$ in Eq.~(\ref{eq:2.5}) can be simplified by introducing $m$ copies of $\phi$ and integrating out the $\zeta$ field, which gives
\begin{equation}
Z_{m}=\lim_{\alpha\rightarrow 0^{+}}\int\mathcal{D}\phi_{a}\exp\left\{-\frac{1}{k_BT}\sum_{a=1}^{m}\mathcal{H}[\phi_{a}]-\frac{\alpha}{2mk_BT}\sum_{1\leq a< b\leq m}^{m}\int d^{3}x\left[\phi_{a}(x)-\phi_{b}(x)\right]^2\right\}, \label{eq:2.6}
\end{equation}
where $a,b$ are replica indices.
Equation~(\ref{eq:2.6}) has the same form as the replicated partition function for a random system with quenched disorder \cite{MezardParisibook}, although here we are interested in the physical limit corresponding to $m=1$. 

To characterize the physical states of the system, we introduce the (renormalized) correlation functions $\mathcal{G}(q)=\left\langle\phi_a(q)\phi_a(-q)\right\rangle$ and $\mathcal{F}(q)=\left\langle\phi_a(q)\phi_b(-q)\right\rangle_{a\neq b}$. $\mathcal{G}(q)$ is the normal physical correlation function of the system, whereas $\mathcal{F}(q)$ measures the correlation between different replicas. It has been shown that $\mathcal{F}(q)$ is equivalent to the long-time correlation function in the conventional mode-coupling approach \cite{Monasson95,Westfahl2001, KirkpatrickJPhysA89}. At high temperatures the system is ergodic, and in the limit $\alpha\rightarrow0^+$ different replicas are not coupled; thus, $\mathcal{F}(q)=0$. When ergodicity is spontaneously broken, different replicas become coupled even in the limit $\alpha\rightarrow0^+$, and $\mathcal{F}(q)\neq0$.  
Using $\mathcal{F}(q)$ as the order parameter for ergodicity breaking, we can define the onset of glassiness $T_A$ as the temperature when there first appears a solution with $\mathcal{F}(q)\neq0$. $T_A$ defined in this way coincides with the dynamic-transition temperature in mean-field--spin-glass models characterized by the appearance of drastically slow dynamic relaxation \cite{Monasson95, Westfahl2001}.  

To obtain the replica free energy defined by Eq.~(\ref{eq:2.5}), we adopt the self-energy approach \cite{Jackiw} and express the effective potential $F_m$ as a functional of bare and renormalized correlation functions
\begin{equation}
F_m\left[\mathbf{G}\right]=\frac{1}{m}\left\{\frac{1}{2}\mathtt{Tr}\ln \mathbf{G}^{-1}+\frac{1}{2}\mathtt{Tr}\left(\mathbf{D}^{-1}\mathbf{G}\right)-\Gamma_2[\mathbf{G}]\right\}. \label{eq:4.6}
\end{equation}
\end{widetext}
 Here $\mathbf{D}$ and $\mathbf{G}$ are bare and renormalized correlation functions, with
\begin{equation}
\mathbf{G}=\left(\mathcal{G}-\mathcal{F}\right)\mathbf{I}+\mathcal{F}\mathbf{E},\label{eq:4.5}
\end{equation}
where $E_{ab}=1$ and $I_{ab}=\delta_{ab}$. [Henceforth we use bold face uppercase letters ($\mathbf{G}$, $\mathbf{D}$, $\bm{\Sigma}$) for the matrices of functions in the replica space, plain uppercase letters with subscript indices ($G_{ab}$, $D_{ab}$, etc.) for the matrix elements. $\mathcal{G}$ and $\mathcal{F}$ are reserved for the renormalized diagonal (physical) and cross-replica correlation functions respectively.] 

$D_{ab}$ is the replica-symmetric bare correlation function, $D_{ab}(q)=g(q)\delta_{ab}$, with $g(q)$ given from Eq.~(\ref{eq:2.12})
\[q_m^{-2}g(q)^{-1}=\frac{1}{4}\left(\frac{q^{2}}{q_{m}^{2}}-1\right)^{2}+\tau_{0}.\] 
The self-energy functions $\Sigma_{ab}$ are defined by
\begin{equation}
\bm{\Sigma}=\mathbf{G}^{-1}-\mathbf{D}^{-1}\label{eq:4.3}
\end{equation}
and obtained through variation of $F_m$:
\begin{eqnarray}
\bm{\Sigma}&=&-\frac{2\delta \Gamma_2[\mathbf{G}]}{\delta\mathbf{G}}.\label{eq:4.4}
\end{eqnarray} 
$\Gamma_2[\mathbf{G}]$ contains all two-particle-irreducible (2PI) diagrams, which is evaluated perturbatively. Detailed calculations are given in Appendix~\ref{Appendix:A}.

Taking the inverse of $\mathbf{G}$ defined in Eq.~(\ref{eq:4.5}), we find that the self-energy from Eq.~(\ref{eq:4.3}) takes the form 
\begin{equation}
\Sigma_{ab}=(\Sigma_{\mathcal{G}}-\Sigma_{\mathcal{F}})\delta_{ab}+\Sigma_{\mathcal{F}},\label{eq:4.9}
\end{equation}
where
\begin{eqnarray}
\Sigma_{\mathcal{G}}(q)&=&\mathcal{G}(q)^{-1}-g(q)^{-1},\label{eq:4.10}\\
\Sigma_{\mathcal{F}}(q)&=&\mathcal{G}(q)^{-1}-\frac{1}{\mathcal{G}(q)-\mathcal{F}(q)}.\label{eq:4.11}
\end{eqnarray}

Assuming that the momentum dependence of self-energy functions $\Sigma_{\mathcal{G}}(q)$ and $\Sigma_{\mathcal{F}}(q)$ is negligible compared with $g(q)$, we can approximate the renormalized diagonal correlation function as 
\begin{widetext}
\begin{equation}
q_m^{-2}\mathcal{G}(q)^{-1}  \approx \frac{1}{4} \left(\frac{q^{2}}{q_{m}^{2}}-1\right)^{2}+\tau_0+\Sigma_{\mathcal{G}}(q_m)q_m^{-2}\equiv\frac{1}{4} \left(\frac{q^{2}}{q_{m}^{2}}-1\right)^{2}+r.\label{eq:4.2}
\end{equation}
And the off-diagonal correlation function $\mathcal{F}(q)$ takes the form
\begin{eqnarray}
\mathcal{F}(q) & = &\mathcal{G}(q)-\frac{1}{\mathcal{G}(q)^{-1}-\Sigma_{\mathcal{F}}(q)}\label{eq:4.7}\\
                      & \approx &\frac{q_m^{-2}}{\frac{1}{4} \left(\frac{q^{2}}{q_{m}^{2}}-1\right)^{2}+r}-\frac{q_m^{-2}}{\frac{1}{4} \left(\frac{q^{2}}{q_{m}^{2}}-1\right)^{2}+r-q_m^{-2}\Sigma_\mathcal{F}(q_m)}\\
                     & \equiv &\frac{q_m^{-2}}{\frac{1}{4} \left(\frac{q^{2}}{q_{m}^{2}}-1\right)^{2}+r}-\frac{q_m^{-2}}{\frac{1}{4} \left(\frac{q^{2}}{q_{m}^{2}}-1\right)^{2}+s}.\label{eq:4.12}
\end{eqnarray}

Equations.~(\ref{eq:4.3}) and (\ref{eq:4.4}) give the self-consistent equations for $\mathcal{G}$ and $\mathcal{F}$ (algebraic equations for $r$ and $s$ in our case). Solving these equations we obtain a normal replica-symmetric solution with $r=s$ and a replica-symmetry-broken solution with $r<s$ below the dynamic-transition temperature $T_A$ [corresponding to some $\left(\chi N\right)_A$ in our diblock-copolymer model].

The configurational entropy is obtained from Eqs.~(\ref{eq:2.7}) and (\ref{eq:4.6}) to be
\begin{equation}
\frac{S_{c}}{k_B} = -\frac{1}{2}\int\frac{d^{3}q}{(2\pi)^{3}}\left[\ln\left(1-\frac{\mathcal{F}(q)}{\mathcal{G}(q)}\right)+\frac{\mathcal{F}(q)}{\mathcal{G}(q)}\right]-\left.\frac{\partial}{\partial{m}}\left(\frac{\Gamma_{2}}{m}\right)\right|_{m=1}. \label{eq:4.8}
\end{equation} 
\end{widetext}
One indeed finds that $S_c$ becomes extensive below $T_A$ and decreases to zero at $T=T_K<T_A$; $T_K$ determines the Kauzmann temperature or the thermodynamic glass transition defined above.

\section{\label{sec:3} Results and Discussion}

\subsection{\label{sec:3.A} Glass transition}

Glass transitions in the Coulomb-frustrated-magnet model have been addressed by several groups in recent years \cite{Nussinov1999,Grousson2001b,Grousson2002a,Grousson2002b,Schmalian2000a,Westfahl2001,Reichman2004,Nussinov2004b,Wolynes2004}. 
These studies establish that in this model glass transitions are possible and could be kinetically favored. 
However, all these studies focus on the strong-coupling regime, and except in Ref.~\cite{Wolynes2004}, the asymmetric cubic interaction has been ignored. 
The block-copolymer system we are studying belongs to the same universality class as the Coulomb-frustrated-magnet--both are examples of the Brazovskii model. 
But for long chains our system corresponds to the weak-coupling regime of the Brazovskii model. 
Furthermore, the presence of the cubic interaction, reflecting compositional asymmetry in the copolymer, is the rule rather than exception. 
It has a strong effect on the ODT and the glass transition, as we will discuss in this work. 

\begin{figure}[b]
 \includegraphics[angle=-90,width=8.8cm]{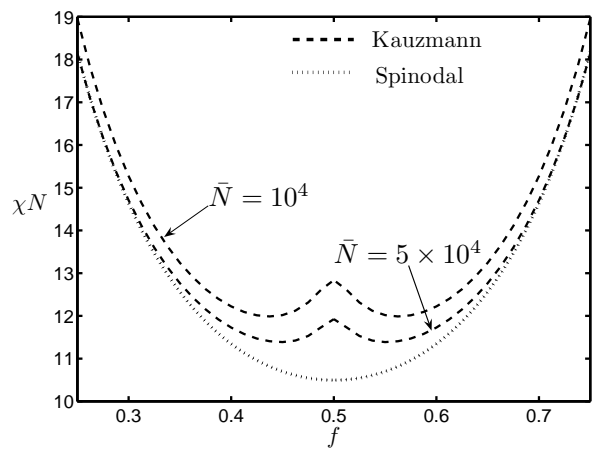}
 \caption{\label{FIG1}Glass transitions in diblock-copolymer melt. Dashed lines are for the Kauzmann temperature and the dotted line for the mean-field spinodal. The upper dashed line is for chain length $\bar{N}=10^4$; the lower one for $\bar{N}=5\times10^4$. Since $T_A$ and $T_K$ are very close, only the $T_K$ transition is shown here.}
\end{figure}

We start with the glass transition. Figure~\ref{FIG1} shows the transition lines for two chain lengths $\bar{N}=10^{4}$ and $\bar{N}=5\times10^{4}$. 
The dotted line represents the mean-field spinodal; dashed lines represent the Kauzmann temperature (or the thermodynamic glass transition temperature \cite{Westfahl2001}) $T_{K}$. The dynamic glass transition temperature $T_A$ is found to be close to the Kauzmann temperature $T_K$ on the scale of this figure in both cases, so we do not present $T_A$ here and only include it in Figs.~2 and 3. In the energy-landscape theory of glass transitions, $T_A$ signals the onset of glassy behavior (e.g., slow dynamics), whereas $T_K$ represents the limit of supercooling below which the system becomes vitrified \cite{Debenedetti00}. (Note that we use the term ``temperature'' even though the phase diagram is presented in terms of the Flory-Huggins interaction parameter $\chi N$; the actual temperature can be determined from the temperature dependence of $\chi N$.) These results show that in diblock-copolymer melts, glass transitions occur at finite temperatures at any chain composition $f$. The narrow gap between $T_A$ and $T_K$ suggests that the system becomes vitrified right after the onset of glassiness. Furthermore, as the chain length increases, both $T_A$  and $T_K$ transitions approach the mean-field spinodal. This latter result is consistent with our anticipation that a large number of inhomogeneous metastable free-energy minima emerge as the system approaches the mean-field spinodal.

Figure~\ref{FIG1} also shows the full crossover from nearly symmetric copolymer, whose glass transitions are dominated by the quartic coupling, 
to highly asymmetric copolymer dominated by the cubic coupling. (We again remind the reader that the results for highly asymmetric block compositions should only be taken as qualitatively but not quantitatively valid.)
For symmetric or nearly symmetric copolymer, it is well known \cite{Grinstein81,FredricksonHelfand} that the mean-field spinodal is destroyed by fluctuations and the disordered phase is always locally stable. 
Also the transition from the disordered phase to the lam phase is a first-order transition with rather complicated (and probably slow) kinetics \cite{HohenbergSwift95a, FredricksonBinder}. 
Therefore a deep quench without annealing can result in the trapping of the system in randomly microphase-separated structures; these structures represent the glassy state captured here.
This scenario is consistent with the experimental observations of Bates \textit{et~al}.~\cite{Bates90a}, where they studied the mechanical properties of three different samples: a rapidly quenched sample, a slowly supercooled sample, and a shear-oriented sample. 
By analogy to molecular liquids, these three samples can be likened to the glassy state, the supercooled-liquid state, and the ordered crystalline state, respectively. 
The quenched sample in this study exhibits solidlike responses at low frequencies while the supercooled sample has typical liquidlike responses. 

We notice that in going from symmetric to asymmetric compositions on either side, the transition lines exhibit a minimum. This is attributed to the crossover from the quartic-coupling dominant to the cubic-coupling dominant regime. As we will discuss later, the cubic term considerably stabilizes the glassy state and enlarges the region of glassy state in the phase diagram. This results in the initial drop of $\chi N$ values at the transitions as $f$ deviates from 0.5 . 

For very asymmetric copolymers, mean-field theory predicts a first-order transition into ordered spherical phases [face-centered cubic (fcc) or bcc] at $\chi N$ smaller than the mean-field spinodal $(\chi N)_S$ \cite{MatsenBatesRev96}. 
However, experiments show that between the featureless disordered phase and the ordered bcc phase, there exists an intervening disordered-micelle state \cite{Register94,Register96,Schwab96,Kim99Mac,Han00a,Hashimoto03a,Han03a,Lodge02,Semenov89Mac, Lodge01}. 
In a self-consistent-field calculation, it was shown \cite{Wang05Mac} that the micelles are formed via a thermally activated process, with a free-energy barrier vanishing at the mean-field spinodal. 
Therefore, if the system is quickly quenched to the vicinity of the spinodal, micelles will proliferate all over the sample; the jamming of these micelles causes their translational diffusion to be so slow that long-range order cannot be developed. 

The interplay between the glass transition and the ODT is complicated. We will present some tentative results in the next subsection. But here we simply note that, in contrast to the symmetric case where glass transitions occur at $\chi N>(\chi N)_S$ (or below the mean-field spinodal temperature), for asymmetric copolymer glass transitions can occur at $\chi N<(\chi N)_S$ (or above the mean-field spinodal temperature). We attribute this to the fact that different arrangements of micelles could generate a large number of metastable states, which significantly stabilize the glassy state.

\begin{figure}
 \includegraphics[angle=-90,width=8.8cm]{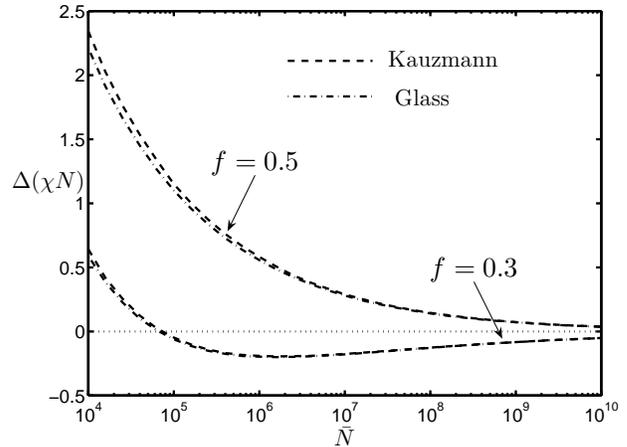}
\caption{\label{FIG2}Chain-length dependence of glass transitions. Plot of $\Delta\left(\chi N\right)\equiv\chi{N}-(\chi N)_S$ against $\bar{N}$. Dashed lines are for the Kauzmann temperature, and dash-dotted lines are for the onset of glassiness. Upper dashed and dash-dotted lines are for the symmetric copolymer, and lower ones for the asymmetric copolymer with $f=0.3$.}
\end{figure}

As highlighted in Eq.~(\ref{eq:2.11}), the chain length $\bar{N}$ controls the magnitude of nonlinear fluctuations and, hence, the deviation from mean-field behavior that is recovered in the limit $\bar{N}\rightarrow\infty$. 
Figure~\ref{FIG2} shows the chain-length dependence of the glass transition temperatures [measured by $(\chi N)_A$ and $(\chi N)_K$, respectively] relative to the mean-field spinodal for symmetric and asymmetric ($f=0.3$) copolymer melts. 
It is clear that in both cases the glass transitions (both $T_A$ and $T_K$) approach the mean-field spinodal as $\bar{N}$ goes to infinity (though from different directions in symmetric and asymmetric cases), implying that in this limit the mean-field spinodal is true stability limit of the disordered phase (with respect to either ordered or randomly phase-separated structures). In other words, the mean-field spinodal is ultimately responsible for the appearance of random structures, thus is the mean-field signature for the glass transition. This general conclusion is likely to be universal to the class of models with continuous degeneracy in the ground states, such as the Brazovskii model (see Ref.~\cite{Nussinov2004b} for other models of the same class). We note that the connection between the spinodal and the glass transition was also implied in an earlier study by Bagchi and co-workers \cite{RicePRB83} of the Lennard-Jones liquid, where the authors conjectured that the liquid spinodal corresponds to the state of random close packing in an equivalent hard-sphere system.

We now discuss the chain-length dependence of the gap between glass transitions and the spinodal, $(\chi N)_A-(\chi N)_S$ and $(\chi N)_K-(\chi N)_S$.  
By a simple scaling analysis given in Appendix \ref{Appendix:B}, we find that both should scale as $\bar{N}^{-0.3}$ for symmetric copolymer. 
This is consistent with the analysis of Wu \textit{et~al}.~\cite{Wolynes2004} if we substitute the $\bar{N}$ dependence of the parameters into their scaling relation. 
Our more accurate numerical calculations confirm this result, as shown in Fig.~\ref{FIG3}, where the first-order transition into lam phase is also included for comparison. 

For asymmetric copolymers, the results are more complicated: for short chains, both $(\chi N)_{A}$ and $(\chi N)_{K}$ are larger than the spinodal value $(\chi N)_{S}$; as $\bar{N}$ increases, the transition lines first shift downward below the spinodal line, which indicates a possible crossover; then, for even larger $\bar{N}$, the transitions gradually approach the spinodal and eventually collapse to the spinodal as $\bar{N}\rightarrow\infty$. 
In this latter limit, we find, using the scaling analysis outlined in Appendix \ref{Appendix:B}, that $(\chi N)_S-(\chi N)_{A,K}\sim \bar{N}^{-1/4}$. 
We attribute the nonmonotonic dependence on $\bar{N}$ to the crossover from the quartic-coupling dominant to cubic-coupling dominant regime as the chain length increases.
Generally for asymmetric copolymers, quartic coupling dominates for short chains and the glass-transition lines are located above the spinodal (or below the spinodal temperature); for long chains, the opposite holds. 

We close this subsection with a brief discussion of the dynamics of the system. 
For the Coulomb-frustrated-magnet model, by invoking the entropy-droplet picture, Wolynes and co-workers predicted \cite{Wolynes2000a, Westfahl2001} that the system should exhibit relaxations similar to fragile liquids \cite{Angell95Science}, characterized by the Vogel-Fulcher behavior, with a relaxation time $\tau\propto\exp[A/(T-T_{K})]$ between $T_{A}$ and $T_{K}$, and diverging at the Kauzmann temperature $T_K$. 
This prediction was disputed by Geissler and Reichman \cite{Reichman2004}, who performed dynamic Monte Carlo simulations of the Brazovskii model without the cubic interaction. They found that as the system approaches the glass-transition temperature predicted by the mode-coupling theory, the relaxation time indeed increases dramatically, but does not show characteristics of fragile liquids. 
Schmalian, Wu, and Wolynes subsequently argued \cite{WolynesComment} that the failure to find the expected dynamic behavior could be a result of the mode-coupling approximation which overestimates the transition temperature. 
Here we note that the simulations by Geissler \textit{et~al.} were performed at temperatures above the ODT temperature, but our calculations show that in the absence of the cubic interaction, the onset of glassiness always occurs below the ODT temperature, at least in the weak-coupling regime. 
Therefore simulations at lower temperatures (below the ODT temperature) are necessary in order to elucidate the dynamic behavior of this model.

For block-copolymer melts, the situation is even more complicated. 
The Hamiltonian given by Eq.~(\ref{eq:2.2}) is a coarse-grained description of the system that focuses on the physics at length scales comparable to or larger than the size of the polymer. 
Therefore we expect the validity of our analysis to be limited to this range of length scales. 
The configurational entropy $S_c$ defined above only measures the number of configurations of chain aggregrates in the locally phase-separated structures, but does not account for different chain conformations within each aggregrate. 
Indeed, above the glass-transition temperature of the \textit{monomer}, polymer chains remain liquidlike even though the system acquires solidlike behavior at the microstructural scale (at high frequencies). Chain diffusion also provides an additional mechanism for relaxations.
Therefore, to accurately describe the dynamic relaxations in block-copolymer melts, one has to consider relaxations both at the microstructural scale and of individual chains.  

\subsection{\label{sec:3.B}Glass-transition vs order-disorder transition}
Our analysis in the previous subsection shows that the glass transition is possible in diblock-copolymer melts and is related to the underlying mean-field spinodal of the disordered phase, which is responsible for the proliferation of inhomogeneous metastable states. However, the ODT also occurs in the neighborhood of the spinodal. Thus a full understanding of the glass transition in this system must address the relationship between these two transitions. 

In molecular liquids the glass transition always takes place in the supercooled state--below the melting (freezing) temperature. 
However, in diblock-copolymer melts, the structural entities forming the random structures are themselves molecular aggregates formed through self-assembly, the number and size of which depend on the temperature of the system. Therefore the relationship between the glass transition and the ODT is not obvious.

Microphase transitions in block-copolymer systems have been extensively studied, both theoretically and experimentally \cite{BatesFredricksonRev91, Hamley98}. 
In previous theories only the Hartree term arising from the quartic interaction was retained; fluctuations due to the cubic interaction were ignored. 
However, our analysis in the previous subsection shows that fluctuations due to the cubic interaction play an essential role in the glass transitions in asymmetric diblock copolymers, at least for long chains
\footnote{In the system of short chains the situation is ambiguous as the higher-order diagrammatic terms neglected in our analysis could become important.}. 
Moreover, as discussed in Sec.~\ref{sec:2.B}, the leading cubic diagram is of the same order in $\bar{N}$ as the Hartree term and their numerical magnitudes are comparable. Therefore we need to include fluctuations due to the cubic term in our studies to have a consistent comparison between the ODT and the glass transition.

In this subsection, we compare the transitions into the ordered phase and into the glassy state.  We have chosen to study symmetric ($f=0.5$) and highly asymmetric $f\sim0.1$ copolymers, as our perturbative methods are better controlled in these two limits (dominated by the quartic and cubic nonlinear interactions, respectively). 

\begin{figure}[t]
 \includegraphics[angle=-90,width=8.8cm]{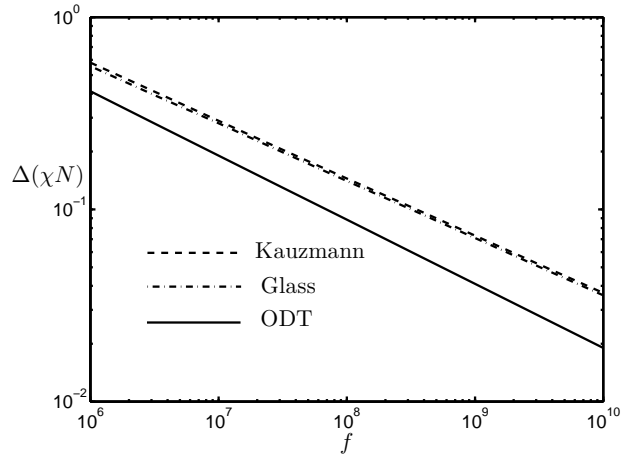}
\caption{\label{FIG3} Glass transition vs ODT in symmetric copolymer melts. Dashed and dash-dotted lines have the same meanings as in Fig.~\ref{FIG2}; the solid line represents the ODT into the lam phase.}
\end{figure}

Figure~\ref{FIG3} shows the chain length dependence of the transitions for symmetric copolymers; it can be considered as a generalized phase diagram.  
The solid line delineates the equilibrium phase boundary between the disordered phase and the ordered lam phase.  
For a given $\bar{N}$, as temperature decreases [$\Delta (\chi N)$ increases], the equilibrium state of the system will change from the disordered phase through a weakly first-order transition to the lam phase.  
However, since the nucleation kinetics is generally slow and complicated \cite{FredricksonBinder, HohenbergSwift95a}, if the system is supercooled to avoid the nucleation of lam phase, the system will remain in a metastable disordered state below the ODT temperature.  
Upon further cooling to the temperature $T_A$ shown as the dash-dotted line, the system enters the glassy regime.  
The region bounded by this line and the Kauzmann line (the dashed line) defines the dynamic range within which glass transition can take place \cite{Monasson95,Schmalian2000a}.  
Although the lam phase has the lowest free energy at low temperatures, once a system is supercooled below $T_K$, it becomes essentially frozen and incapable of reaching the more stable lam state. The narrow gap between the onset of glassiness and the Kauzmann temperature implies that the glass transition in block copolymer melts will be fairly sharp.

In symmetric copolymer melts we observe the scaling of $(\chi N)_{ODT}-(\chi N)_S \sim \bar{N}^{-1/3}$ as predicted in Ref.~\cite{FredricksonHelfand}. 
For the onset of glassiness, $(\chi N)_A-(\chi N)_S$ scales as $\bar{N}^{-0.3}$, which agrees well with our approximate scaling analysis given in Appendix \ref{Appendix:B}.
Our results show that for symmetric copolymers, the ODT always occurs before the glass transitions (i.e., at temperatures above the glass transitions). 
While one might argue that this conclusion could be due to the particular choice of diagrams in our perturbative calculation, we find that this scaling with $\bar{N}$ remains unchanged when a different approximation scheme, the self-consistent-screening approximation, is used \footnote{C.-Z.~Zhang and Z.-G.~Wang, unpublished}. 
In addition, our results are also consistent with the local-field calculations by Wu \textit{et~al}.~\cite{Wolynes2004}, as will be discussed later in this subsection. 
Since $(\chi N)_{ODT}-(\chi N)_S$ decays more rapidly with $\bar{N}$ than both $(\chi N)_A-(\chi N)_S$ and $(\chi N)_K-(\chi N)_S$, for sufficiently large $\bar{N}$, we always have $(\chi N)_{ODT} < (\chi N)_{A,K}$. 
Therefore, at least in the long-chain limit, our conclusion that the glass transition occurs below the ODT temperature should be valid, regardless of the approximations in the calculation. 

\begin{figure}[b]
 \subfigure[$\bar{N}=10^{7}$]{\includegraphics[angle=-90,width=8.8cm]{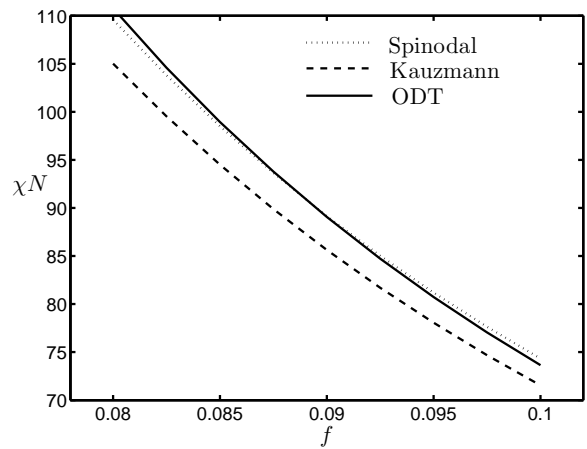}}\hspace{.5in}
 \subfigure[$\bar{N}=10^{8}$]{\includegraphics[angle=-90,width=8.8cm]{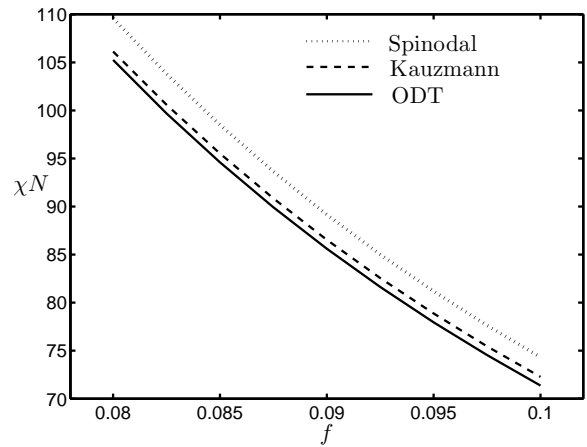}}
\caption{\label{FIG4} Glass transition vs. ODT in very asymmetric copolymer. (a) $\bar{N}=10^{7}$, from above: mean-field spinodal (dotted line), ODT (solid line), and Kauzmann temperature (dashed line). (b) $\bar{N}=10^{8}$, from above: mean-field spinodal (dotted line), Kauzmann temperature (dashed line), and ODT (solid line).}
\end{figure}

Figure~\ref{FIG4} shows various transitions for highly asymmetric copolymers around $f=0.1$. Here again $T_A$ is not shown as it is very close to $T_K$ on the scale of the figures.
In the case of $\bar{N}=10^{7}$, the glass-transition lines are located below the ODT, i.e., the glass transition temperatures are above the ODT temperature. 
In other words, glass transitions can precede the ordering transition into the bcc phase.
This unusual behavior is quite different from what happens in molecular fluids, where the glass transition always occurs below the freezing (ordering) temperature. 
In the case of longer chains with $\bar{N}=10^{8}$, the ODT occurs before the glass transitions; this is the expected behavior in the asymptotic limit $\bar{N}\rightarrow\infty$, since in this limit the glass-transition lines approach the mean-field spinodal whereas the ODT into the bcc phase takes place at a finite distance below the spinodal \cite{MatsenBatesRev96}. 

The chain-length dependence of the glass transitions relative to the ODT for asymmetric diblock copolymers is qualitatively similar to the critical micelle temperature in the same system. 
It is shown \cite{Wang05Mac} that disordered micelles can appear in large numbers before the ordering transition only for not-too-long chains; for very long chains, the ODT will set in before the disordered micelles reach a considerable concentration, essentially precluding the disordered micelles from being a distinct intervening phase between the featureless disordered state and the ordered (fcc or bcc) phases. 
Since micelles are likely to be the structural entities in the glassy asymmetric copolymer melts, the connection between the micelle formation and the glass transition is worth further investigation.

As discussed in Sec.~\ref{sec:3.A}, the cubic interaction stabilizes the glassy state. 
We attribute this stabilizing effect to the additional complexities in the configurational space caused by the cubic term. 
This effect is closely related to the effect of the cubic interaction on the ODT. 
Theoretical analysis shows that the presence of the cubic term can considerably reduce the free energy of ordered microstructures with three fold symmetries. 
This is consistent with the fact that there are more stable ordered phases in asymmetric diblock copolymers. 
If we visualize the random structures as polycrystals with local but no long-range order, then the increased variety of mesophase structures will increase the complexity in the configuration space \footnote{For a discussion on the possible differences between polycrystalline phases and structural glasses in this context, see Ref.~\cite{Wolynes2004}.}, which can explain the stabilization of the glassy state in asymmetric copolymers.

As a final technical point, we compare our treatment of the cubic term with that in Ref.~\cite{Wolynes2004}. 
There the authors used a local-field approximation, in which a momentum-independent self-energy is solved variationally by mapping the Brazovskii Hamiltonian [as given by Eq.~(\ref{eq:2.11})] to a reference nonlinear but local Hamiltonian. 
Within this approximation, it was found that the cubic interaction considerably stabilizes the glassy state and the glass transition can occur at temperatures above the mean-field spinodal temperature; these results coincide with ours. 
However, for certain choices of parameters in the weak-coupling regime, this local-field treatment could result in a nonmonotonic relation between the correlation length and the temperature. We believe this unphysical behavior is probably due to overestimating the fluctuation effects due to the cubic term in their treatment.

\section{\label{sec:5}Conclusions}

To conclude, using the thermodynamic replica formalism we have shown that at low temperatures, diblock-copolymer melts can exist as randomly microphase-separated structures, in addition to the thermodynamically stable periodic structures. 
This transition is essentially a glass transition in which the supercooled liquid gradually gets vitrified. 
We have identified the temperature range over which this glass transition can occur, which is bordered by the onset of glassiness (or the dynamic glass transition) temperature from above and the Kauzmann (thermodynamic glass transition) temperature from below. 
For symmetric diblock copolymers, the glass transition takes place below the temperature of the ODT into the lam phase. 
However, for asymmetric diblock copolymers, the glass transition can precede the ordering transition, which is an unusual feature that probably reflects the self-assembly nature of the system. 
This study leads us to naturally identify the quenched samples of block copolymers in some previous experimental works as the glassy state of the system. 
Given the slow phase transition kinetics in copolymer systems, we expect such glassy structures to be quite common in these systems without externally imposed aligning fields.

As in any theories on polymer mixtures \cite{FredricksonHelfand, Wang02a}, the scaled degree of polymerization, $\bar{N}$, serves as a Ginzburg parameter which allows us to systematically examine the approach to mean-field behavior as $\bar{N} \rightarrow \infty$.  
An important conclusion is that in the limit of infinitely long chains, the glass transitions collapse to the mean-field spinodal, suggesting that the mean-field spinodal is ultimately responsible for the proliferation of inhomogeneous free-energy minima and can be used as the mean-field signature for the glass transition. 

That a glass transition occurs at the mean-field spinodal in the limit of $\bar{N} \to \infty$ can also be understood using the following dynamical argument.  Since the Hamiltonian has an overall factor of ${\bar{N}}^{1/2}$, in the mean-field approximation, we expect the free energy barriers between the metastable states to be proportional to this factor.  For very long chains, these barriers can be very large.  Since proliferation of the metastable minima appears at the spinodal \cite{Nussinov2004b}, upon a quench below the spinodal, the system will first go to these metastable states with overwhelming probability because of their large number, and transitions from these metastable states should be very slow. Note that it is the barriers from these metastable states to the (more stable) ordered phases and between the metastable states themselves, rather than the nucleation barrier from the uniform disordered phase to the ordered phases, that are relevant to the glass transition. Hence, for example, in symmetric diblock copolymers, even though the transition from the disordered to lamellar phase approaches second order in the limit ${\bar{N}} \rightarrow \infty$ (where the nucleation barrier vanishes \cite{FredricksonBinder}), our theory predicts a glass transition that coincides with the ODT, which is the spinodal in this limit.

Studying diblock-copolymer melt as a specific example of the Brazovskii model, we find that the cubic interaction significantly increases the stability of the glassy state as well as the bcc phase, and causes qualitative changes in the scaling relations with the chain length. 
We conjecture that this stabilizing effect is due to increased configurational complexity as a result of more free-energy minima due to the presence of the cubic term.
 
\begin{acknowledgments}
This work was supported by the National Science Foundation through the Center for the Science and Engineering of Materials at Caltech. 
\end{acknowledgments}

\appendix

\begin{widetext}

\section{\label{Appendix:A} Perturbative expansion of the effective potential with broken symmetries}

In this appendix we present the details of our perturbative calculation of the free energy defined in Eq.~(\ref{eq:2.5}). The general expansion of the effective potential for a system with broken symmetry was derived in Ref.~\cite{Jackiw}. Here we omit the details of that derivation but give the result
\begin{equation}
\Gamma[\varphi, \mathbf{G}]=I[\varphi]+\frac{1}{2}\mathtt{Tr}\ln\mathbf{G}^{-1}+\frac{1}{2}\mathtt{Tr}\left(\mathbf{D}^{-1}\mathbf{G}\right)-\Gamma_2[\mathbf{G}; \Delta H].\label{eq:A.0}
\end{equation}
Here $\varphi$ is the order parameter in the ordered phase, $I[\varphi]$ is the mean-field free energy (in our case the Leibler free energy), $\Delta H[\psi; \varphi]$ is the shifted Hamiltonian [see Eq.~(\ref{eq:3.6})], $\mathbf{D}$ is the shifted bare propagator defined as
\begin{equation}
D_{ab}(q)=\frac{\delta^2\Delta H[\varphi]}{\delta\varphi_a(q)\delta\varphi_b(-q)},\label{eq:A.10}
\end{equation}
and $\mathbf{G}$ is the renormalized propagator.
As noted before, we reserve boldface uppercase letters for matrices of correlation functions and use the corresponding plain ones when referring to the matrix element.
The second term ($\mathtt{Tr}\ln$) in Eq.~(\ref{eq:A.0}) is the one-loop correction and the last term, $\Gamma_2$ contains higher-order corrections, including all 2PI diagrams generated by the vertices in the shifted Hamiltonian $\Delta H$ with the renormalized propagator $\mathbf{G}$. The third term ensures the consistency of the expansion in terms of the renormalized propagator. 

It has been shown in Ref.~\cite{Jackiw} that $\Gamma[\varphi, \mathbf{G}]$ as defined in Eq.~(\ref{eq:A.0}) is stationary with respect to both $\varphi$ and $\mathbf{G}$. Therefore one can derive the self-energy equations through a variation of Eq.~(\ref{eq:A.0}), which gives
\begin{equation}
\frac{\delta\Gamma[\mathbf{G}]}{\delta\mathbf{G}}= 0 \Rightarrow \bm{\Sigma}= \mathbf{G}^{-1}-\mathbf{D}^{-1}=-\frac{2\delta\Gamma_2[\mathbf{G}]}{\delta\mathbf{G}}.\label{eq:A.11}
\end{equation}
In the field-theory description of diblock-copolymer melts [Eq.~(\ref{eq:2.11})], $\bar{N}^{-1/2}$ serves as a smallness parameter, which enables a straightforward loop expansion for $\Gamma_2[\mathbf{G}]$. To the leading two-loop order (one-loop order in the self-energy), one has three terms
\begin{subequations}
\label{eq:A.1-4}
\begin{eqnarray}
\Gamma_2^{(1)} & = & -\frac{\lambda}{8\bar{N}}\sum_{a}\int\frac{d^{3}q_{1}d^{3}q_{2}}{(2\pi)^{6}}{G}_{aa}(q_{1}){G}_{aa}(q_{2}) \label{eq:A.1},\\
\Gamma_2^{(2)} & = &  \frac{\eta^{2}}{12\bar{N}}\sum_{a,b}\int\frac{d^{3}q_{1}d^{3}q_{2}}{(2\pi)^{6}}{G}_{ab}(q_{1}){G}_{ab}(q_{2}){G}_{ab}(-q_{1}-q_{2})\label{eq:A.2},\\
\Gamma_2^{(3)} & = & \frac{\lambda^{2}}{12\bar{N}}\sum_{a,b}\int\frac{d^3q_1d^3q_2d^3q_3}{(2\pi)^9}\varphi_a(q_1){G}_{ab}(-q_2){G}_{ab}(-q_3){G}_{ab}(q_1+q_2+q_3)\varphi_b(-q_1),\label{eq:A.4}
\end{eqnarray} 
corresponding to the diagrams shown in Figs.~\ref{feyn}(a), 5(b), and 5(c) respectively. In the glassy state, translational symmetry breaking does not occur; therefore, $\varphi=0$, $D_{ab}(q)=g(q)\delta_{ab}$, and $\Gamma_2^{(3)}$ vanishes. Note that $\Gamma_2^{(1)}$ is the Hartree term which only generates a momentum-independent self-energy in the diagonal part of $\mathbf{G}$; $\Gamma_2^{(2)}$ generates an off-diagonal self-energy which enables a nontrivial solution with broken replica symmetry. For symmetric copolymer, the cubic coupling is zero; therefore, to find possible solutions with broken replica symmetry we need to include the off-diagonal term of second order,
\begin{equation}
\Gamma_{2}^{(4)} =\frac{\lambda^{2}}{48\bar{N}^{3/2}}\sum_{a,b}\int\frac{d^{3}q_{1}d^{3}q_{2}d^{3}q_{3}}{(2\pi)^{9}}{G}_{ab}(q_{1}){G}_{ab}(q_{2}){G}_{ab}(q_{3}){G}_{ab}(-q_{1}-q_{2}-q_{3}),\label{eq:A.3}
\end{equation} 
\end{subequations}
corresponding to the three-loop diagram as shown in Fig.~\ref{feyn}(d). To study the crossover from very asymmetric to symmetric copolymer, we keep $\Gamma_{2}^{(4)}$ in the off-diagonal renormalization for asymmetric copolymer as well. 

From Eqs.~(\ref{eq:A.11}) and (\ref{eq:A.1-4}) we obtain the self-energy
\begin{equation}
\Sigma_{ab}(k) = \frac{\lambda}{2\bar{N}^{1/2}}\int\frac{d^{3}q}{(2\pi)^{3}}{G}_{aa}(q)\delta_{ab}
-\frac{\eta^{2}}{2\bar{N}^{1/2}}\int\frac{d^3{q}}{(2\pi)^3}G_{ab}(q)G_{ab}(k-q)-\frac{\lambda^{2}}{6\bar{N}}\int\frac{d^{3}qd^3p}{(2\pi)^{6}}{G}_{ab}(q)G_{ab}(p)G_{ab}(k-p-q).
\end{equation}
The three terms on the right-hand side corresponding to Fig.~\ref{feyn}(e), 5(f), and 5(g), respectively.

\unitlength=1mm  

\begin{figure}
\subfigure[]{
  \begin{fmffile}{feyn3}
  \begin{fmfgraph}(25,30)
  \fmfpen{thick}
  \fmfleft{i}
  \fmfright{o}
  \fmf{phantom}{i,v,o}
  \fmf{plain,left=90,tension=0.8}{v,v}
  \fmf{plain,left=-90,tension=0.8}{v,v}
  \fmfdot{v}
  \end{fmfgraph}
  \end{fmffile}
}
 \hspace{.1in}
 \subfigure[]{
  \begin{fmffile}{feyn4}
  \begin{fmfgraph}(30,30)
  \fmfpen{thick}
  \fmfleft{i}
  \fmfright{o}
  \fmf{phantom}{i,v1}
  \fmf{phantom}{v2,o}
  \fmf{plain,left,tension=0.3}{v1,v2,v1}\fmffreeze
  \fmf{plain}{v1,v2}
  \fmfdotn{v}{2}
  \end{fmfgraph}
  \end{fmffile}
 }
\hspace{.1in}
\subfigure[]{
 \begin{fmffile}{feyn6}
\begin{fmfgraph}(30,30)
\fmfpen{thick}
\fmfleft{i}
\fmfright{o}
\fmf{wiggly}{i,v1}
\fmf{plain,left,tension=0.3}{v1,v2,v1}
\fmf{wiggly}{v2,o}\fmffreeze
\fmf{plain}{v1,v2}
\fmfdotn{v}{2}
\end{fmfgraph}
\end{fmffile}
}
\hspace{.1in}
\subfigure[]{
\begin{fmffile}{feyn7}
\begin{fmfgraph}(30,30)
\fmfpen{thick}
\fmfleft{i}
\fmfright{o}
\fmf{phantom}{i,v1}\fmf{phantom}{v2,o}
\fmf{plain,left=0.5,tension=0.3}{v1,v2,v1}
\fmffreeze
\fmf{plain,left=1}{v1,v2,v1}
\fmfdotn{v}{2}
\end{fmfgraph}
\end{fmffile}
}
\hspace{.1in}
\subfigure[]{
\begin{fmffile}{feyn1} 
  \begin{fmfgraph}(20,25)
  \fmfpen{thick}
  \fmfleft{i}
  \fmfright{o}
  \fmf{plain,tension=0.45}{i,v,v,o}
  \fmfdot{v}
  \end{fmfgraph}
\end{fmffile}}
\hspace{.1in} 
 \subfigure[]{
 \begin{fmffile}{feyn2}
  \begin{fmfgraph}(30,25) 
  \fmfpen{thick}
  \fmfleft{i}
  \fmfright{o}
  \fmf{plain}{i,v1}
  \fmf{plain,left=0.8,tension=0.2}{v1,v2,v1}
  \fmf{plain}{v2,o}
  \fmfdotn{v}{2}
  \end{fmfgraph}
  \end{fmffile}
 }
\hspace{.1in} 
 \subfigure[]{
 \begin{fmffile}{feyn5}
  \begin{fmfgraph}(30,25)
  \fmfpen{thick}
  \fmfleft{i}
  \fmfright{o}
  \fmf{plain}{i,v1}
  \fmf{plain,left=0.8,tension=0.2}{v1,v2,v1}
  \fmf{plain}{v2,o}\fmffreeze
  \fmf{plain}{v1,v2}
  \fmfdotn{v}{2}
  \end{fmfgraph}
 \end{fmffile} 
 } 
\caption{
\label{feyn}Feynman diagrams: (a)--(d) Loop diagrams in $\Gamma_2$. (e)--(g) Self-energy diagrams. In the diagrams thick solid lines represent the renormalized propagator $\mathbf{G}$ and wiggly lines represent the external leg of the order parameter $\varphi(q)$. We use a slightly different perturbative expansion for the diagonal renormalization, as explained in Appendix \ref{Appendix:C}.
}
\end{figure}
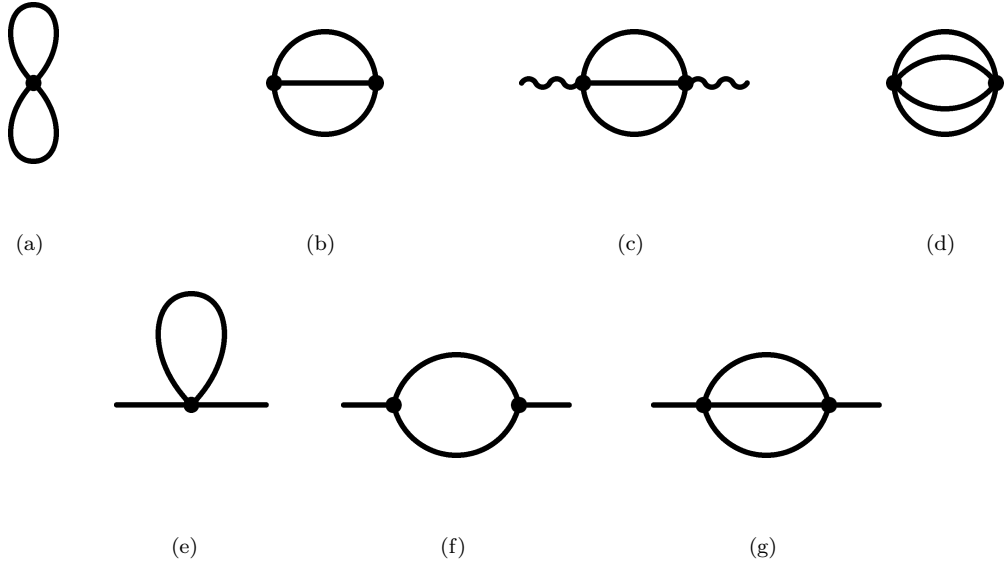

Under the one-step--replica-symmetry-breaking (1-RSB) ansatz, $G_{ab}(q)=\left[\mathcal{G}(q)-\mathcal{F}(q)\right]\delta_{ab}+\mathcal{F}(q)$, and
$\Gamma_{2}$ has three terms from Eqs.~(\ref{eq:A.1}), (\ref{eq:A.2}), and (\ref{eq:A.3})
\begin{subequations}
\label{eq:A.12-14}
\begin{eqnarray}
\Gamma_{2}^{(1)} =& & -\frac{m\lambda}{8\bar{N}}\left(\int\frac{d^{3}q}{(2\pi)^{3}}\mathcal{G}(q)\right)^{2},\label{eq:A.12}\\
\Gamma_{2}^{(2)} =& &\frac{\eta^2}{12\bar{N}}\left[m\int\frac{d^{3}q_{1}d^{3}q_{2}}{(2\pi)^{6}}\mathcal{G}(q_{1})\mathcal{G}(q_{2})\mathcal{G}(-q_{1}-q_{2})+m(m-1)\int\frac{d^{3}q_{1}d^{3}q_{2}}{(2\pi)^{6}}\mathcal{F}(q_{1})\mathcal{F}(q_{2})\mathcal{F}(-q_{1}-q_{2})\right],\label{eq:A.13}\\
\Gamma_{2}^{(4)} =& &\frac{\lambda^2}{48\bar{N}^{3/2}}\left[m\int\frac{d^{3}q_{1}d^{3}q_{2}d^{3}q_{3}}{(2\pi)^{9}}\mathcal{G}(q_{1})\mathcal{G}(q_{2})\mathcal{G}(q_{3})\mathcal{G}(-q_{1}-q_{2}-q_{3})\right.\nonumber\\
             && + \left.m(m-1)\int\frac{d^{3}q_{1}d^{3}q_{2}d^{3}q_{3}}{(2\pi)^{9}}\mathcal{F}(q_{1})\mathcal{F}(q_{2})\mathcal{F}(q_{3})\mathcal{F}(-q_{1}-q_{2}-q_{3})\right].\label{eq:A.14}
\end{eqnarray}
\end{subequations}
 
Using the polarization functions $\Pi_{ab}(k)$ defined as
\begin{subequations}
\begin{eqnarray}
\Pi_{ab}(k) & = &\int\frac{d^{3}q}{(2\pi)^{3}}{G}_{ab}(q){G}_{ba}(k+q)\nonumber\\
            & = &\left[\Pi_{\mathcal{G}}(k)-\Pi_{\mathcal{F}}(k)\right]\delta_{ab}+\mathcal{F}(k), \nonumber\\ 
\Pi_{\mathcal{G}}(k) & = & \int\frac{d^{3}q}{(2\pi)^{3}}\mathcal{G}(q)\mathcal{G}(k+q),\\
\Pi_{\mathcal{F}}(k) & = & \int\frac{d^{3}q}{(2\pi)^{3}}\mathcal{F}(q)\mathcal{F}(k+q).
\end{eqnarray}
\end{subequations}
we can rewrite the self-energy functions (after taking $m=1$) as
\begin{subequations}
\begin{eqnarray}
\Sigma_{ab}& = & \left(\Sigma_{\mathcal{G}}-\Sigma_{\mathcal{F}}\right)\delta_{ab}+\Sigma_{\mathcal{F}},\nonumber \\
\Sigma_{\mathcal{G}}(k) & = & \frac{\lambda}{2\bar{N}^{1/2}}\int\frac{d^{3}q}{(2\pi)^{3}}\mathcal{G}(q)-\frac{\eta^{2}}{2\bar{N}^{1/2}}\Pi_{\mathcal{G}}(k)-\frac{\lambda^{2}}{6\bar{N}}\int\frac{d^{3}q}{(2\pi)^{3}}\mathcal{G}(q)\Pi_{\mathcal{G}}(q+k)\label{eq:A.6},\\
\Sigma_{\mathcal{F}}(k) & = & -\frac{\eta^{2}}{2\bar{N}^{1/2}}\Pi_{\mathcal{F}}(k)-\frac{\lambda^{2}}{6\bar{N}}\int\frac{d^{3}q}{(2\pi)^{3}}\mathcal{F}(q)\Pi_{\mathcal{F}}(q+k).\label{eq:A.7}
\end{eqnarray}
\end{subequations}

The configurational entropy is obtained from Eqs.~(\ref{eq:4.8}) and (\ref{eq:A.12-14}) to be 
\begin{subequations}
\label{eq:A.8-9}
\begin{eqnarray}
S_{c}       & = & \frac{1}{T}\left.\frac{\partial{F_{m}}}{\partial{m}}\right|_{m=1}=S_{c}^{(0)}+S_{c}^{(1)}, \nonumber\\
S_{c}^{(0)} & = & -\frac{1}{2\bar{N}^{1/2}}\int\frac{d^{3}q}{(2\pi)^{3}}\left[\ln\left(1-\frac{\mathcal{F}(q)}{\mathcal{G}(q)}\right)+\frac{\mathcal{F}(q)}{\mathcal{G}(q)}\right],\label{eq:A.8}\\
S_{c}^{(1)} & = & -\frac{\eta^{2}}{12\bar{N}}\int\frac{d^{3}q}{(2\pi)^{3}}\Pi_{\mathcal{F}}(q)\mathcal{F}(-q)-\frac{\lambda^{2}}{48\bar{N}^{3/2}}\int\frac{d^{3}q}{(2\pi)^{3}}\Pi_{\mathcal{F}}(q)\Pi_{\mathcal{F}}(-q).\label{eq:A.9}
\end{eqnarray}
\end{subequations}
We have set $k_BT=1$ in above derivations; the configurational entropy is given in unit of $k_B$ per unit volume.

\section{\label{Appendix:C} Order-disorder transition}

In this appendix, we present our calculation of the ODT in diblock-copolymer melts. Our approach is different from the Brazovskii approximation \cite{Brazovskii75, FredricksonHelfand}. 

Following the derivation in Appendix \ref{Appendix:A}, we expand the effective potential to two-loop order and keep only the diagonal terms in Eqs.~(\ref{eq:A.1-4}):
\begin{eqnarray}
\Gamma[\bar\phi=\varphi]&=&F_{L}(\varphi)+\frac{1}{2\bar{N}^{1/2}}\mathtt{Tr}\ln\mathcal{G}^{-1}+\frac{1}{2\bar{N}^{1/2}}\mathtt{Tr}(D^{-1}\mathcal{G})+\frac{\lambda}{8\bar{N}}\left[\int\frac{d^{3}{q}}{(2\pi)^3}\mathcal{G}(q)\right]^{2}\nonumber\\
                                        & &-\frac{\eta^{2}}{12\bar{N}}\int\frac{d^{3}q}{(2\pi)^{3}}\mathcal{G}(q)\Pi_{\mathcal{G}}(-q)-\frac{\lambda^{2}}{12\bar{N}}\int\frac{d^3q_1d^3q_2}{(2\pi)^6}\varphi(q_1)\mathcal{G}(-q_2)\Pi_{\mathcal{G}}(q_1+q_2)\varphi(-q_1)\label{eq:C.1},
\end{eqnarray}
where $F_{L}(\varphi)$ is the Leibler free energy for the ordered phase, $D(q)$ is the shifted bare propagator as given in Eq.~(\ref{eq:3.2}). The Hartree approximation (similar to the Brazovskii approximation) amounts to keeping only the first four terms of Eq.~(\ref{eq:C.1}), which can be justified by a renormalization-group argument \cite{Shankar94}. The central idea is the following: since near the critical temperature the dominant fluctuations are those with wave numbers close to $q_m$ at which the propagator is maximized, one can decompose the spherical shell into small ``patches'' and rewrite the order parameter into $n$ components, each corresponding to one patch. In this way one can rewrite the original Hamiltonian [Eq.~(\ref{eq:2.11})] as an $n$-vector model. At the critical point, $n$ goes to infinity and the Hartree approximation becomes exact. Therefore we may replace $\mathcal{G}$ by the Hartree approximation $\mathcal{G}_H$ as defined in Eq.~(\ref{eq:3.3})
\[\mathcal{G}_H(k)^{-1}=D(k)^{-1}+\frac{\lambda}{2\bar{N}^{1/2}}\int\frac{d^{3}q}{(2\pi)^{3}}\mathcal{G}_H(q).\]
This gives the first three terms in Eq.~(\ref{eq:3.7}).

However, here we want to study the correction due to the cubic coupling; thus, we want to include the leading-order diagram from the cubic interaction in the effective potential, as shown in Fig.~\ref{feyn}(b). It can be shown that in the corresponding $n$-vector model as mentioned above, the Hartree term [Fig.~\ref{feyn}(a)] is of order $O(n)$ and this correction term [Fig.~\ref{feyn}(b)] is of order $O\left(n^{1/2}\right)$. In our numerical calculations we find these two terms to be comparable for the temperature range we are interested in. By a similar argument the last term in Eq.~(\ref{eq:C.1}) is of order $O(1)$ and ignored in our calculation (the numerical value is indeed small compared with the other one-loop diagrams because of the weak first-order nature of the transition). To summarize, the free energy is given by Eq.~(\ref{eq:C.1}) with the last term dropped, as is Eq.~(\ref{eq:3.4}). 

To find the ODT temperature, the free energy is minimized numerically with respect to the magnitude of density wave $a$ as given in Eq.~(\ref{eq:3.1}) and the ODT occurs when the free energy of the ordered phase equals the free energy of the disordered phase.

The physical correlation function is given by Eq.~(\ref{eq:3.5}) and the corresponding self-energy is 
\begin{equation}
\Sigma_{\mathcal{G}}(k)= \frac{\lambda}{2\bar{N}^{1/2}}\int\frac{d^{3}q}{(2\pi)^{3}}\mathcal{G}_H(q)-\frac{\eta^{2}}{6\bar{N}^{1/2}}\int\frac{d^3q}{(2\pi)^3}\mathcal{G}_H(q)\mathcal{G}_H(q+k).\label{eq:A.6'}
\end{equation}
This renormalization scheme includes two parts, the first corresponding to a simple Hartree approximation and the second incorporating fluctuations from the cubic interaction using the Hartree-renormalized propagator, which is schematically shown in the following diagrammatic equations: 
\begin{fmffile}{feyneq1}
\begin{eqnarray}
\parbox{22mm}{\begin{fmfgraph}(20,20)
  \fmfpen{thin}
  \fmfleft{i}
  \fmfright{o}
  \fmf{dbl_plain}{i,o}
  \end{fmfgraph}
} & = &
\parbox{22mm}{\begin{fmfgraph}(20,20)
  \fmfpen{thin}
  \fmfleft{i}
  \fmfright{o}
  \fmf{plain,width=5}{i,o}
  \end{fmfgraph}
}- 
\frac{\lambda}{2}\parbox{22mm}{\begin{fmfgraph}(20,20)
  \fmfpen{thin}
  \fmfleft{i}
  \fmfright{o}
  \fmf{plain,width=7}{i,v}
  \fmf{dbl_plain}{v,o}
  \fmf{dbl_plain,left=-90, tension=0.8}{v,v}
  \fmfdot{v}
  \end{fmfgraph}
},\nonumber\\
&&\\
\parbox{22mm}{\begin{fmfgraph}(20,20)
  \fmfpen{thick}
  \fmfleft{i}
  \fmfright{o}
  \fmf{plain}{i,o}
\end{fmfgraph}
} & = &
\parbox{22mm}{\begin{fmfgraph}(20,20)
  \fmfleft{i}
  \fmfright{o}
  \fmf{dbl_plain}{i,o}
  \end{fmfgraph}
} +
\frac{\eta^2}{2}\parbox{22mm}{\begin{fmfgraph}(20,20) 
  \fmfleft{i}
  \fmfright{o}
  \fmf{dbl_plain}{i,v1}
  \fmf{dbl_plain,left=0.8,tension=0.3}{v1,v2,v1}
  \fmf{plain,width=20}{v2,o}
  \fmfdotn{v}{2}
  \end{fmfgraph}
},\nonumber
\end{eqnarray}
\end{fmffile}
where thin lines represent the bare propagator $g(q)$, double lines represent the Hartree-renormalized propagator $\mathcal{G}_H(q)$, and thick lines represent the physical propagator $\mathcal{G}(q)$. Equation~(\ref{eq:A.6'}) modifies the self-energy equation (\ref{eq:A.6}) we derived using a straighforward loop expansion in Appendix \ref{Appendix:A}.  One can verify that this self-energy equation does not have the unphysical nonmonotonic relation between temperature (manifested through $\chi N$ in $\tau_0$) and correlation length [manifested in $r$ in $\mathcal{G}(q)$] which occurs in a naive loop expansion, and this renormalization scheme indeed gives consistent result in the known limits; e.g. when $(\chi N)_S-\chi N\gg 1$ it reduces to the loop expansion and when $(\chi N)_S-\chi N\sim 0$ it gives the leading-order terms in the $1/n$ expansion. 

\section{\label{Appendix:B} Approximate solution of the glass transition} 

In this last appendix we provide an approximate solution of the self-consistent equations obtained in Appendix \ref{Appendix:A}. The diagonal and off-diagonal self-energy equations are shown in the following diagrammatic equations:
\begin{fmffile}{feyneq2}
\begin{eqnarray}
\parbox{22mm}{\begin{fmfgraph*}(20,20)
\fmfpen{thick}
\fmfleft{i}
\fmfright{o}
\fmf{plain}{i,v,o}
\fmfv{decor.shape=circle,decor.filled=empty,decor.size=.6w,label=$\Sigma_\mathcal{G}$,label.dist=-0.1w}{v}
\end{fmfgraph*}
}
& = &
\frac{\lambda}{2}\parbox{22mm}{\begin{fmfgraph}(15,20)
  \fmfleft{i}
  \fmfright{o}
  \fmf{plain}{i,v}
  \fmf{plain}{v,o}
  \fmf{dbl_plain,left=-90, tension=0.5}{v,v}
  \fmfdot{v}
  \end{fmfgraph}
}
-\frac{\eta^2}{2}\parbox{22mm}{\begin{fmfgraph}(20,20) 
  \fmfleft{i}
  \fmfright{o}
  \fmf{plain}{i,v1}
  \fmf{dbl_plain,left=0.8,tension=0.2}{v1,v2,v1}
  \fmf{plain}{v2,o}
  \fmfdotn{v}{2}
  \end{fmfgraph}
},\\
\parbox{22mm}{\begin{fmfgraph*}(20,20)
\fmfpen{thick}
\fmfleft{i}
\fmfright{o}
\fmf{dashes}{i,v,o}
\fmfv{decor.shape=circle,decor.filled=shaded,decor.size=.6w,label=$\Sigma_\mathcal{F}$, label.dist=-0.1w}{v}
\end{fmfgraph*}
} &  = & 
-\frac{\eta^2}{2} \parbox{22mm}{\begin{fmfgraph}(20,20) 
  \fmfpen{thick}
  \fmfleft{i}
  \fmfright{o}
  \fmf{dashes}{i,v1}
  \fmf{dashes,left=0.8,tension=0.2}{v1,v2,v1}
  \fmf{dashes}{v2,o}
  \fmfdotn{v}{2}
  \end{fmfgraph}
}-
\frac{\lambda^2}{6} \parbox{22mm}{\begin{fmfgraph}(20,20)
  \fmfpen{thick}
  \fmfleft{i}
  \fmfright{o}
  \fmf{dashes}{i,v1}
  \fmf{dashes,left=0.8,tension=0.2}{v1,v2,v1}
  \fmf{dashes}{v2,o}\fmffreeze
  \fmf{dashes}{v1,v2}
  \fmfdotn{v}{2}
  \end{fmfgraph}
},
\end{eqnarray}
\end{fmffile}
\end{widetext}
where dashed lines represent the renormalized off-diagonal propagator $\mathcal{F}$; thick lines and double lines represent the renormalized diagonal propagator $\mathcal{G}$ and the Hartree-renormalized propagator, respectively, the same as before.

From Eqs.~(\ref{eq:4.2}) and (\ref{eq:4.12}) the renormalized propagators $\mathcal{G}$ and $\mathcal{F}$ are given by
\begin{equation}
\mathcal{G}(q) = \frac{4q_{m}^{-2}}{\left(q^2/q_{m}^{2}-1\right)^{2}+4r},\tag{28$'$}
\end{equation}
\begin{equation}
\mathcal{F}(q) = \frac{4q_{m}^{-2}}{\left(q^2/q_{m}^{2}-1\right)^{2}+4r}-\frac{4q_{m}^{-2}}{\left(q^2/q_{m}^{2}-1\right)^{2}+4s}\tag{31$'$}.
\end{equation}
When $r$,$s$ are small, the polarization functions can be approximated as
\begin{subequations}
\begin{eqnarray}
\Pi_{\mathcal{G}}(k) & \simeq & \frac{1}{4kr},\label{eq:B.4}\\
\Pi_{\mathcal{F}}(k) & \simeq & \frac{1}{4k}\left(\frac{1}{\sqrt{r}}-\frac{1}{\sqrt{s}}\right)^{2},\label{eq:B.5}
\end{eqnarray}
\end{subequations}
for $0<\left|k\right|<2$ and zero elsewhere. These are verified numerically and work well for $r,s$ not too large $(\lesssim0.1)$. 
The diagrammatic terms in our calculations are found to be
\begin{subequations}
\begin{equation}
\int\frac{d^3q}{(2\pi)^3}\mathcal{G}(q)\approx\frac{q_m}{2\pi\sqrt{r}},
\end{equation}
\begin{equation}
\left.\int\frac{d^{3}q}{(2\pi)^{3}}\mathcal{G}(k-q)\Pi_{\mathcal{G}}(q)\right|_{k=q_{m}} \approx \frac{1}{8\pi r\sqrt{r}},\label{eq:B.6}
\end{equation}
\begin{equation}
\left.\int\frac{d^{3}q}{(2\pi)^{3}}\mathcal{F}(k-q)\Pi_{\mathcal{F}}(q)\right|_{k=q_{m}} \approx  \frac{1}{8\pi}\left(\frac{1}{\sqrt{r}}-\frac{1}{\sqrt{s}}\right)^{3},\label{eq:B.7}
\end{equation} 
\begin{equation}
\int\frac{d^{3}q}{(2\pi)^{3}}\mathcal{F}(-q)\Pi_{\mathcal{F}}(q) \approx \frac{1}{8\pi}\left(\frac{1}{\sqrt{r}}-\frac{1}{\sqrt{s}}\right)^{3},\label{eq:B.8}
\end{equation}
\begin{equation}
\int\frac{d^{3}q}{(2\pi)^{3}}\Pi_{\mathcal{F}}(-q)\Pi_{\mathcal{F}}(q) \approx \frac{q_m}{16\pi^{2}}\left(\frac{1}{\sqrt{r}}-\frac{1}{\sqrt{s}}\right)^{4}.\label{eq:B.9}
\end{equation}
\end{subequations}

From Eqs.~(B3) and (28$'$) we have the following equations for $r$
\begin{eqnarray}
\tau_0 & = & \tau-\frac{\lambda}{4\pi\bar{N}^{1/2} q_{m}\sqrt{\tau}},\nonumber\\
&&\label{eq:B.10}\\
r      & = & \tau-\frac{\eta^{2}}{8\pi\bar{N}^{1/2} q_{m}^{3}\tau}.\nonumber
\end{eqnarray}
And from Eqs.~(C2) and (31$'$) we have
\begin{equation}
s-r = \frac{\lambda^{2}}{48\pi\bar{N} q_{m}^{2}}\left(\frac{1}{\sqrt{r}}-\frac{1}{\sqrt{s}}\right)^{3}+\frac{\eta^{2}}{8\bar{N}^{1/2}q_{m}^{3}}\left(\frac{1}{\sqrt{r}}-\frac{1}{\sqrt{s}}\right)^{2}.\label{eq:B.11}
\end{equation}

Let us look at Eq.~(\ref{eq:B.11}) first. Defining $t\equiv\sqrt{r/s}$, Eq.~(\ref{eq:B.11}) becomes
\begin{equation}
t^{-2}-1 = \frac{\lambda^{2}(1-t)^{3}}{48\pi\bar{N} q_{m}^{2}r^{5/2}}+\frac{\eta^{2}(1-t)^{2}}{8\bar{N}^{1/2}q_{m}^{3}r^{2}}.\label{eq:B.12}
\end{equation}
This equation always has a replica-symmetric solution $t=1$ ($\mathcal{F}=0$). Here we are seeking a replica-symmetry-broken solution with $t<1$. Defining the dimensionless parameters
\begin{eqnarray*}
A & \equiv & \frac{\lambda^{2}}{48\pi\bar{N} q_{m}^{2}r^{5/2}},\\
B & \equiv & \frac{\eta^{2}}{8\bar{N}^{1/2}q_{m}^{3}r^{2}},
\end{eqnarray*}
Eq.~(\ref{eq:B.12}) becomes
\begin{equation}
\frac{A(1-t)^{2}t^2}{1+t}+\frac{B(1-t)t^2}{1+t}=1.\tag{C7$'$}\label{eq:B.13}
\end{equation}
Numerical calculations show that when both $A$ and $B$ are non-negative and either $A>23.66$ or $B>11.09$, there is always a solution $0<t^{\ast}<1$. For symmetric copolymer and very asymmetric copolymer, respectively, these inequalities result in the criteria
\begin{eqnarray}
r & \lesssim & \left(\frac{\lambda^{2}}{\bar{N}q_{m}^{2}}\right)^{2/5}\sim\bar{N}^{-2/5},\label{eq:B.14}\\
r & \lesssim & \left(\frac{\eta^{2}}{\bar{N}^{1/2}q_{m}^{3}}\right)^{1/2}\sim\bar{N}^{-1/4}.\label{eq:B.15}
\end{eqnarray}
And the resulted scaling relations for $\tau_{0}$ [$\propto (\chi N)_S-\chi N$] for symmetric and asymmetric copolymers are, respectively,
\begin{eqnarray}
\tau_0 & \sim & -\bar{N}^{-0.3},\label{eq:B.16}\\
\tau_0 & \sim & \bar{N}^{-1/4}.\label{eq:B.17}
\end{eqnarray}
These have been verified by our numerical calculations.

Finally we look at the configurational entropy and the Kauzmann temperature.
The configurational entropy is given in Eqs.~(\ref{eq:A.8-9}) and found to be
\begin{subequations}
\begin{equation}
S_{c}^{(0)}  = \frac{q_m^{3}\sqrt{r}}{4\pi\bar{N}^{1/2} t}(1-t)^{2},\label{eq:B.18}
\end{equation}
\begin{equation}
S_{c}^{(1)} \approx -\frac{\eta^{2}}{96\pi\bar{N}r^{3/2}}(1-t)^3-\frac{\lambda^{2}q_{m}}{768\pi^{2}\bar{N}^{3/2}r^{2}}(1-t)^4.\label{eq:B.19}
\end{equation}
\end{subequations}
Thus for symmetric copolymer ($\eta=0$), the Kauzmann transition is located at
\begin{eqnarray}
r \sim \bar{N}^{-2/5},\label{eq:B.23}\\
\tau_{0} \sim -\bar{N}^{-0.3}.\label{eq:B.24}
\end{eqnarray}
And for very asymmetric copolymer ($\eta/q_m^{3/2}\gg\lambda/q_m$),
\begin{eqnarray}
r \sim \bar{N}^{-1/4},\label{eq:B.25}\\
\tau_{0} \sim \bar{N}^{-1/4}.\label{eq:B.26}
\end{eqnarray}
                      
\section{\label{Appendix:D}Relationship between the pinned free energy $F[\zeta]$ and the free-energy landscape of the original Hamiltonian $H[\phi]$}

Here we explicitly show that the free energy $F[\zeta]$ defined in Eq.~(\ref{eq:2.4}) captures the metastable free-energy minima of the Hamiltonian $H[\phi]$ as defined in Eq.~(\ref{eq:2.11}).
First we rewrite Eq.~(\ref{eq:2.4}) as
\[ F[\zeta]=-\ln\int\mathcal{D}\phi\exp\left(- \mathcal{H'}[\phi]+\alpha\zeta*\phi-\frac{\alpha}{2}\zeta*\zeta\right),\]
where $*$ is a shorthand notation for integration and
\begin{equation}
H'[\phi]=H[\phi]+\frac{\alpha}{2}\phi*\phi.
\end{equation}
We then define the generating functional of the perturbed Hamiltonian  $H'[\phi]$,
\begin{equation}
W[J=\alpha\zeta]=-\ln\int\mathcal{D}\phi\exp\left(- \mathcal{H'}[\phi]+J*\phi\right).
\end{equation}
The effective potential $\Gamma'[\varphi]$ of the Hamiltonian  $H'[\phi]$ is obtained as the Legendre transform of $W[J]$. Thus we have
\begin{eqnarray}
\left.\frac{\delta W[J]}{\delta  J}\right|_{J=\alpha\zeta}&=&-\varphi,\label{eq:D.3}\\
\Gamma'[\varphi]&=&W[J]+J*\varphi,\\
\left.\frac{\delta \Gamma'[\varphi]}{\delta  \varphi}\right|_{\varphi}&=&J.\label{eq:D.5}
\end{eqnarray}
Now $W[J]$ is related to $F[\zeta]$ by
\begin{equation}
W[J=\alpha\zeta]=F[\zeta]-\frac{\alpha}{2}\zeta*\zeta,
\end{equation}
so that for any $\zeta^\ast$ that minimizes $F[\zeta]$, we have
\begin{equation}
\left.\frac{\delta F[\zeta]}{\delta  \zeta}\right|_{\zeta=\zeta^\ast}=\alpha\left.\frac{\delta W[J]}{\delta  J}\right|_{J=\alpha\zeta^\ast}+\alpha\zeta^\ast=0,
\label{eq:D.6}
\end{equation}
that is,
\begin{equation}
\zeta^\ast=-\left.\frac{\delta W[J]}{\delta  J}\right|_{J=\alpha\zeta^\ast}=\varphi^\ast.
\label{eq:D.7}
\end{equation}
Equation~(\ref{eq:D.7}) holds for any positive $\alpha$, including in particular the limit $\alpha\rightarrow0^+$.
In the limit of $\alpha\rightarrow0^+$, $H'[\phi]$ approaches $H[\phi]$  and $\Gamma'[\varphi]$ approaches $\Gamma[\varphi]$, the effective  potential of the original Hamiltonian $H[\phi]$.  Also $J=\alpha\zeta\rightarrow0$,  from  Eq.~(\ref{eq:D.5}), $\varphi^\ast$ becomes a minimum of  $\Gamma[\varphi]$.  This, together with Eq.~(\ref{eq:D.7}), shows that  the minima of $F[\zeta]$ coincide with the minima of the effective  potential $\Gamma[\varphi]$ of the orginal Hamiltonian in the limit  $\alpha\rightarrow0^+$. 


\end{document}